\documentclass[aps,preprint,pre,superscriptaddress,amsmath,amssymb]{revtex4-1}
\usepackage{mathrsfs}
\usepackage{upgreek}

\usepackage{ifthen,calc}

\def\longrightharpoonup{\relbar\joinrel\rightharpoonup}
\def\longleftharpoondown{\leftharpoondown\joinrel\relbar}

\def\longrightleftharpoons{
  \mathop{
    \vcenter{
      \hbox{
      \ooalign{
        \raise1pt\hbox{$\longrightharpoonup\joinrel$}\crcr
	  \lower1pt\hbox{$\longleftharpoondown\joinrel$}
	  }
      }
    }
  }
}

\newcommand{\aref}[1]{App.~\ref{app:#1}}

\newcommand{\alabel}[1]{\label{app:#1}}

\newcommand{\tref}[1]{Table.~\ref{table:#1}}

\newcommand{\tlabel}[1]{\label{table:#1}}
\newcommand{\fref}[1]{Fig.~\ref{fig:#1}}

\newcommand{\flabel}[1]{\label{fig:#1}}
\newcommand{\eref}[1]{Eq.~\ref{eqn:#1}}
\newcommand{\erefs}[1]{Eqs.~\ref{eqn:#1}}
\newcommand{\erefstwo}[2]{Eqs.~\ref{eqn:#1}~and~\ref{eqn:#2}}
\newcommand{\erefsthree}[3]{Eqs.~\ref{eqn:#1}, ~\ref{eqn:#2}~and~\ref{eqn:#3}}

\newcommand{\elabel}[1]{\label{eqn:#1}}

\newcommand{\beq}{\begin{equation}}
\newcommand{\beqs}{\begin{subequations}}
\newcommand{\beqnn}{\begin{equation*}}
\newcommand{\beqn}{\begin{eqnarray}}
\newcommand{\beqnnn}{\begin{eqnarray*}}
\newcommand{\bml}{\begin{multline}}
\newcommand{\eeq}{\end{equation}}
\newcommand{\eeqs}{\end{subequations}}
\newcommand{\eeqnn}{\end{equation*}}
\newcommand{\eeqn}{\end{eqnarray}}
\newcommand{\eeqnnn}{\end{eqnarray*}}
\newcommand{\eml}{\end{multline}}
\newcommand{\pd}{\partial}

\newcommand{\avg}[1]{\langle{#1}\rangle}

\newcommand{\br}[1]{\left ( #1\right )}

\newcommand{\pdiff}[2]{\frac{\pd #1}{\pd #2}}

\usepackage{tikz}
\usepackage{pifont}

\newcommand*{\FSfont}[1]{}

\definecolor{Dark}{gray}{0.2}
\definecolor{MedDark}{gray}{0.4}
\definecolor{Medium}{gray}{0.6}
\definecolor{Light}{gray}{0.8}
\makeatletter
\DeclareRobustCommand\ltseries
{\not@math@alphabet\ltseries\relax
\fontseries\ltdefault\selectfont}
\newcommand{\ltdefault}{l}
\DeclareTextFontCommand{\textlt}{\ltseries}
\DeclareRobustCommand\hbseries
{\not@math@alphabet\hbseries\relax
\fontseries\hbdefault\selectfont}
\newcommand{\hbdefault}{hb}
\DeclareTextFontCommand{\texthb}{\hbseries}
\makeatother

\begin{document}

\title{The lower bound on the precision of transcriptional regulation}

\author{Joris \surname{Paijmans}}\affiliation{FOM Institute AMOLF, Science Park 104, 1098 XG Amsterdam, The
Netherlands}
\author{Pieter Rein \surname{ten Wolde}}\affiliation{FOM Institute AMOLF, Science Park 104, 1098 XG Amsterdam,
The Netherlands}

\begin{abstract}
  The diffusive arrival of transcription factors at the promoter sites
  on the DNA sets a lower bound on how accurately a cell can regulate
  its protein levels. Using results from the literature on
  diffusion-influenced reactions, we derive an analytical expression
  for the lower bound on the precision of transcriptional regulation.
  In our theory, transcription factors can perform multiple rounds of
  1D diffusion along the DNA and 3D diffusion in the cytoplasm before
  binding to the promoter. Comparing our expression for the lower bound on the precision
  against results from Green's Function Reaction Dynamics simulations shows that the
  theory is highly accurate under biologically relevant
  conditions. Our results demonstrate that, to an excellent
  approximation, the promoter switches between the
  transcription-factor bound and unbound state in a Markovian fashion. 
  This remains true even in the presence of sliding, i.e. with 1D diffusion along the
  DNA. This has two important implications: (1) minimizing the noise
  in the promoter state is equivalent to minimizing the search time of
  transcription factors for their promoters; (2) the complicated
  dynamics of 3D diffusion in the cytoplasm and 1D diffusion along the
  DNA can be captured in a well-stirred model by renormalizing the
  promoter association and dissociation rates, making it possible to
  efficiently simulate the promoter dynamics using Gillespie
  simulations. Based on the recent experimental observation that
  sliding can speed up the promoter search by a factor of 4, our theory
  predicts that sliding can enhance the precision of transcriptional
  regulation by a factor of 2.
\end{abstract}

\maketitle

\section{INTRODUCTION}
Biological cells regulate their protein levels by stimulating or
repressing the expression of genes via the binding of transcription
factors (TFs) to the regulatory sequences on the DNA called promoters. 
The fluctuations in the
state of the promoter, switching between `on' and `off' due to the
binding and unbinding of transcription factors, will 
propagate to the protein levels downstream. Because there are only
very few transcription factors present in a cell and because they have
to find their target site via a diffusive trajectory, these
fluctuations are substantial. Furthermore, in contrast to what has been
assumed before \cite{Hippel1989}, the binding of the TFs to their
target is not diffusion limited \cite{Hammar2012a}. This is likely to
enhance the fluctuations in the promoter state even further.

The level of transcription is set by the fraction of time
the promoter is in the 'on' state. This fraction, in turn, is controlled 
by the TF concentration. But how well can the cell infer the TF
concentration from the strongly fluctuating promoter occupancy? The
diffusion and the limited affinity of the TF for the promoter puts a
fundamental limit on how precise gene expression can be regulated.
In turn, this puts a lower bound on the noise in gene expression.

Indeed, in a computational study by Van Zon et al. \cite{VanZon2006},
it was found that the diffusive arrival of TFs at the promoter is a
major source of noise in gene expression. In their model, however, the
promoter was represented as a sphere, and it was assumed that the
transcription factors move by normal 3D diffusion on all length
scales. However, it is now commonly believed that transcription
factors find their promoter via a combination of 1D diffusion along
the DNA and 3D diffusion in the cytoplasm
\cite{Riggs1972,Richter1974,Hippel1989,Halford2004,Hu2006,Elf2007,Lomholt:2009vj,Li2009,Hammar2012a}.

Recently, it has been studied theoretically how deviations of the TFs
transport from classical Brownian motion affects noise in gene
expression \cite{Tkacik2009,Tamari2011,Kafri2012}. On length scales
larger than the sliding distance, the transport process is
essentially 3D diffusion, but on length scales smaller than the
sliding distance, the dynamics is a complicated interplay of 3D
diffusion in the cytoplasm and 1D diffusion on the DNA. This motivated
Tka\v{c}ik and Bialek to study a model in which TFs can move by 3D
diffusion in the bulk, bind reversibly and non-specifically to DNA
near the promoter, move by 1D
diffusion along the DNA to the promoter, to which they can then
bind specifically and reversibly \cite{Tkacik2009}. Tka\v{c}ik and Bialek found that the effect of the 
larger target size on the noise in gene expression, provided by the 1D sliding along the DNA near the
promoter, is largely canceled by the increased temporal correlations
in 1D diffusion. As a result, sliding has, according to their
analysis, only a small effect on the physical limits to the precision
of transcriptional regulation. 

Here we rederive the fundamental bound on the accuracy of
transcriptional regulation. We study the same model as that of
Tka\v{c}ik and Bialek \cite{Tkacik2009}, but analyze it using the
approach of Agmon, Szabo, and coworkers to study
diffusion-influenced reactions \cite{Agmon1990, Kaizu2014}. Apart from
one biologically motivated assumption and one mathematical
approximation, this approach makes it possible to solve this model
exactly. To test our theory, we have  extended Green's Function
Reaction Dynamics \cite{VanZon2005A,VanZon2005,VanZon2006,Takahashi2010},
which is an exact scheme for simulating reaction-diffusion systems at
the particle level, to include 1D diffusion along cylinders. We find
excellent agreement between the predictions of our theory and the
simulation results. 

Our expression for the sensing error differs qualitatively from that
of Tka\v{c}ik and Bialek \cite{Tkacik2009}. Our expression predicts that, 
as the average promoter occupancy  approaches unity, the error diverges. 
This can be understood intuitively by noting that
in this limit newly arriving TFs cannot bind the promoter, and hence
no concentration measurements can be performed. We found the same
result earlier for the binding of ligand to a spherical receptor
\cite{Kaizu2014}. 

The key ingredient that determines the lower bound on the accuracy of
transcriptional regulation is the correlation time of the promoter
state \cite{Berg1977,Bialek2005,Kaizu2014}. The correlation time is a
complex function of the diffusion constants of the TFs in the
cytoplasm and along the DNA, and the rates of non-specific DNA binding
and specific promoter binding. However, we find that, to an excellent
approximation, the promoter correlation time is that of a random
telegraph process, in which the promoter switches between the TF bound
and unbound state with effective rates that are constant in time.  The
reason is that in living cells, the TF concentration is typically low,
{\it i.e.} in the nM range, while the sliding distance and sliding
time are short, $\approx 50 {\rm bp}$ and $<50 {\rm ms}$, respectively
\cite{Elf2007,Hammar2012a}. As a result, even in the presence of
sliding along the DNA, the time a TF spends near the promoter is short
compared to the timescale on which TFs arrive at the promoter from the
bulk, which is on the order of seconds to minutes
\cite{Hammar2012a}. Hence, a TF near the promoter either rapidly binds
the promoter or rapidly escapes into the bulk.  This makes it possible
to integrate out the rapid promoter-TF rebindings and the unsuccessful
TF bulk arrivals, and reduce the many-body, non-Markovian
reaction-diffusion problem to a pair problem in which the TFs
associate with and dissociate from the promoter with rates that are
constant in time. These results underscore our earlier finding that
the complex TF diffusion dynamics with its algebraic distributed waiting times can
be described in a well-stirred model by renormalizing the association
and dissociation rates. Importantly, this implies that this model can 
then be simulated using the Gillespie algorithm\cite{Gillespie1977,VanZon2006,Kaizu2014}. 

One of the most important implications of our observation that the
promoter dynamics can be described by a random telegraph process, is
that minimizing the promoter noise (correlation time) is equivalent to
minimizing the time required for transcription factors to find and bind
the promoter. As pointed out by Tka\v{c}ik and Bialek, the combined system 
of 1D and 3D diffusion tends to have longer correlation times than the system 
with only 3D diffusion \cite{Tkacik2009}. However, the dominant effect is 
that the DNA binding increases the target size which speeds up the rate by 
which TFs find the promoter. Our results show that this
decreases the promoter correlation time, which enhances the precision
of transcriptional regulation, and lowers the noise in gene
expression. This means that the large body of work on how proteins find their targets on the DNA
\cite{Richter1974,Berg1981,Slutsky2004a,Hu2006,Halford2004,Elf2007,Lomholt:2009vj,Li2009,Hammar2012a}
could be used to study how cells can optimize the precision of
transcriptional regulation. Our findings corroborate those of Hammar et
al. \cite{Hammar2012a}: the search time and hence the promoter noise
(correlation time) can be minimized by optimizing the sliding time.
The optimal sliding time depends on the probability that a TF which is
in contact with the promoter will actually bind the promoter rather
than sliding over it.

\section{THEORY}
Following earlier work \cite{Berg1977,Bialek2005,Tkacik2009,Kaizu2014}, we imagine that the
cell infers the average transcription factor concentration $\bar{c}$
from the promoter state $n(t)$ integrated over an integration time
$T$, $n_T = (1/T) \int_0^T n(t) dt$. Here, $n(t)$ is one if at time
$t$ a transcription factor is bound to the promoter, and zero otherwise.  In
the limit that the integration time $T$ is much longer than the
correlation time $\tau_n$ of $n(t)$, the variance in our estimate
$n_T$ of the true mean occupancy $\bar{n}$ is given by
\cite{Berg1977,Kaizu2014}
\begin{equation}
 \left( \delta n \right)^2 \equiv \sigma_{n, T}^2 \simeq \frac{2 \sigma_{n}^2 \,\tau_{n}}{T}=\frac{P_n\br{\omega=0}}{T},\
 \elabel{std_devT}
\end{equation}
where $\sigma_n^2=\avg{n^2}-\avg{n}^2$ is the variance of an
instantaneous measurement,  and
$P_n(\omega)$ and $\hat{C}_n(s)$ are respectively the power spectrum
and the Laplace transform of the auto-correlation function $C_n(t)$ of
$n(t)$.

The uncertainty or
expected error $\delta c$ in the corresponding estimate of the average
concentration $\bar{c}$ is related to the error $\delta n$ in the
estimate of $\bar{n}$ via the gain $d\bar{n}/d\bar{c}$,
\begin{equation}
\delta c = \left| \frac{d \bar{c}}{d\bar{n}} \right| \delta n.
\elabel{gain}
\end{equation}
Since the promoter is a binomial switch, the variance $\sigma_n^2
=\bar{n}(1-\bar{n})$.  Both the average occupancy $\avg{n}=\bar{n}$
and the gain $d\bar{n}/d\bar{c}$ are determined by the input-output
relation $\bar{n}(\bar{c})$ and the average concentration $\bar{c}$,
while the integration time $T$ is assumed given. Hence, to obtain the
error in the concentration estimate, we need to know the promoter
correlation time $\tau_n$.

We note that the above expressions are generic: they apply to all
systems where the concentration is inferred from the binary binding
state of a protein, be it a receptor on the membrane or a
promoter. How the ligand molecules or the transcription factors
diffuse to the receptor or the promoter only enters the problem via the
magnitude of the receptor (promoter) correlation time.

\subsection{Deriving the correlation function and correlation time}
To derive the uncertainty $\delta n$ in our estimate of $\bar{n}$, we
 derive the correlation function for a binary
switching process (see \eref{std_devT}), following Kaizu et
al. \cite{Kaizu2014}. We start with the general
expression for the correlation function of a binary switch
\begin{eqnarray}
 C_n(\tau) & \equiv & \left\langle (n(\tau) - \bar{n})(n(0) - \bar{n}) \right\rangle \\
  	& = & \bar{n} \left( p_{*|*}(\tau)-\bar{n} \right).
 \elabel{corr_func}
\end{eqnarray}
In the second line we introduced the probability that the promoter is
bound at time $\tau$, given that is started in the bound state at
$t=0$. This conditional probability is equal to
\begin{equation}
 p_{*|*}(\tau) = 1 - \mathscr{S}_{\mathrm{rev}}(\tau|*)
\end{equation}
where $\mathscr{S}_{\mathrm{rev}}(\tau|*)$ is the probability that the
promoter is free at time $\tau$, given that it was bound
initially. The promoter can undergo multiple rounds of binding and
unbinding during the time $\tau$. We can describe this reversible
process in terms of an irreversible one via the convolution
\cite{Agmon1990}
\begin{equation}
 \mathscr{S}_{\mathrm{rev}}(t|*) = k_- \int_0^t [1-\mathscr{S}_{\mathrm{rev}}(t'|*)]\mathscr{S}_{\mathrm{rad}}(t-t'|z_0)dt'.
 \elabel{rev_conv}
\end{equation}
The first factor under the integral gives the probability that the
promoter is occupied at time $t'$. Then the transcription factor
dissociates from the promoter with a rate $k_-$ and
is placed in contact with the promoter on the DNA at position $z_0$. The second term
under the integral, $\mathscr{S}_{\mathrm{rad}}(t-t'|z_0)$, gives the
probability that the promoter remains unoccupied from the last
dissociation up to time time $t$.  Integrating over all intermediate
times $t'$, gives us the probability that the promoter is unoccupied
at time $t$.

To solve \eref{rev_conv}, we need the irreversible survival
probability of the promoter,
$\mathscr{S}_{\mathrm{rad}}(t-t'|z_0)$. In general, this quantity
cannot be analytically, since it depends on the history of
binding events 
\cite{Agmon1990,Kaizu2014}. Following \cite{Agmon1990,Kaizu2014}, we
will assume  that after each
promoter-TF dissociation event, the promoter with
the TF at contact is surrounded by an equilibrium
distribution of TFs.  The survival
probability is then given by
\begin{equation}
 \mathscr{S}_{\mathrm{rad}}(t|z_0) \simeq \mathscr{S}_{\mathrm{rad}}(t|{\mathrm{eq}}) S_{\mathrm{rad}}(t|z_0),
 \elabel{Def_Ssi}
\end{equation}
where $\mathscr{S}_{\mathrm{rad}}(t|{\mathrm{eq}})$ is the survival
probability of a promoter which is free initially and is
surrounded by an equilibrium solution of
TFs; $S_{\mathrm{rad}}(t|z_0)$ is the probability that a free promoter
with only a single TF at contact $z_0$ and no other TFs present, is
still unbound at a later time $t$. Below, in sections \ref{sec:Assumptions} and \ref{sec:ValidityAssumptions}, we discuss the validity
of \eref{Def_Ssi} in detail. 

The quantity $\mathscr{S}_{\mathrm{rad}}(t|{\mathrm{eq}})$ can be found by solving the differential equation (\aref{noise:Seq_derived})
\begin{equation}
 \pdiff{\mathscr{S}_{\mathrm{rad}}(t|{\mathrm{eq}})}{t} = - \bar{\xi} \, k_{\mathrm{rad}}(t) \, \mathscr{S}_{\mathrm{rad}}(t|{\mathrm{eq}}).
 \elabel{DE_Seq}
\end{equation}
Here, $k_{\mathrm{rad}}(t)$ is the time-dependent rate coefficient,
and, importantly, $\bar{\xi}$ is the average concentration of TFs on
the DNA, and not the total concentration of TFs. The above equation relates the rate at which TFs 
that were in equilibrium at time $t=0$ bind the
promoter at time $t$, $-\pdiff{\mathscr{S}_{\mathrm{rad}}(t|{\mathrm{eq}})}{t}$,
to the rate at which TFs  bind the
promoter at time $t$ if it is not occupied, $\bar{\xi} \,
k_{\mathrm{rad}}(t)$, times the probability that the promoter is indeed
unoccupied, $\mathscr{S}_{\mathrm{rad}}(t|{\mathrm{eq}})$.  Solving
the equation yields
\begin{equation}
 \mathscr{S}_{\mathrm{rad}}(t|{\mathrm{eq}}) = e^{-\bar{\xi}\int_0^t k_{{\mathrm{rad}}}(t') dt'}.
 \elabel{Seq}
\end{equation}
Because the system obeys detailed balance, we can write $k_{{\mathrm{rad}}}(t)$ \cite{Agmon1990} as 
\begin{equation}
 k_{\mathrm{rad}}(t) = k_+ \, S_{\mathrm{rad}}(t|z_0),
 \elabel{krad}
\end{equation}
where $k_+$ is the intrinsic association rate of the TF when in
contact with the promoter. 

Before deriving the correlation function $C_n(\tau)$ in the Laplace
domain, $\hat{C}_n(s)$, we give a relation which will prove useful. Namely, from
\erefstwo{DE_Seq}{krad}, it is clear that
\begin{eqnarray}
 \pdiff{\mathscr{S}_{\mathrm{rad}}(t|{\mathrm{eq}})}{t} & = & -\bar{\xi} k_+ \, S_{\mathrm{rad}}(t|z_0) \, \mathscr{S}_{\mathrm{rad}}(t|{\mathrm{eq}}) \\
 & = & -\bar{\xi} k_+ \, \mathscr{S}_{\mathrm{rad}}(t|z_0).
 \elabel{SeqSsi}
\end{eqnarray}

To derive $\hat{C}_n(s)$, we first Laplace transform
\eref{rev_conv} and solve it for
$\hat{\mathscr{S}}_{\mathrm{rev}}(s|*)$. By using the Laplace
transformed \erefstwo{corr_func}{SeqSsi} and using that
$k_-\bar{n}=k_+\bar{\xi} (1-\bar{n})$ and $\sigma^2_n = \bar{n}
(1-\bar{n})$, we can express $\hat{C}_n(s)$ as a function of
$\hat{\mathscr{S}}_{\mathrm{rad}}(s|{\mathrm{eq}})$ only (see also
\cite{Kaizu2014}):
\begin{equation}
 \hat{C_n}(s) = \sigma_n^2 \, \frac{\bar{n} \hat{\mathscr{S}}_{\mathrm{rad}}(s|{\mathrm{eq}})}{1-(1-\bar{n})s\hat{\mathscr{S}}_{\mathrm{rad}}(s|{\mathrm{eq}})}.
 \elabel{LapCorr}
\end{equation}
To obtain an analytically closed form for the correlation function, we
require an expression for
$\hat{\mathscr{S}}_{\mathrm{rad}}(s|{\mathrm{eq}})$. We use 
\begin{equation}
 \hat{\mathscr{S}}_{\mathrm{rad}}(s|{\mathrm{eq}}) \simeq \frac{1}{s}  \, \frac{1}{ 1 + \bar{\xi} \hat{k}_{\mathrm{rad}}(s)},
 \elabel{Seq_approx}
\end{equation}
which correctly captures the short- and long-time limit of
$\mathscr{S}_{\mathrm{rad}}(t|{\mathrm{eq}})$ and becomes exact
  for all times in the low concentration limit \cite{Kaizu2014}.
Substituting this approximation into \eref{LapCorr}, we obtain, after simplifying
\begin{equation}
  \hat{C_n}(s) = \sigma_n^2 \, \frac{\bar{n}}{\bar{n} s + k_+ \bar{\xi} \, s \hat{S}_{\mathrm{rad}}(s|z_0)}.
 \elabel{Corr_approx}
\end{equation}
We can find the correlation time by taking the $s \to 0$ limit of the
correlation function in Laplace space (see \eref{std_devT}).
Using that $\bar{n}=k_+\bar{\xi} / (k_+\bar{\xi}+k_-)$, the expression for the correlation time of the promoter state becomes
\begin{equation}
 \tau_{n} = \lim_{s \to 0} \frac{1}{\sigma_n^2} \hat{C}_{n}(s) = \frac{\tau_c}{S_{\mathrm{rad}}(\infty|z_0)}.
 \elabel{CT1}
\end{equation}
Here $\tau_{c} = \left(k_+ \bar{\xi} + k_-\right)^{-1}$ is the
correlation time of the intrinsic switching dynamics, i.e. the
correlation time of the promoter occupancy when the promoter-TF
association is reaction-limited and the effect of diffusion can be
neglected.  Note that in geometries for which the particle always
returns to the starting point, such as in 1D and 2D diffusion
problems, $\lim_{t\to\infty} S_{\mathrm{rad}}(t|z_0)\to0$, such that
the correlation time in \eref{CT1} diverges. In these geometries, the
particle always remains correlated with its starting point, and we are
unable to define a correlation time. However, in the living cell,
transcription factors do not only diffuse along the DNA, but also
in the cytoplasm where memory is lost, yielding a finite correlation time.

In \aref{noise:rates} we show that $S_{\mathrm{rad}}(\infty|z_0)$ can
be related to the intrinsic promoter-TF binding rate $k_+$ 
and the promoter-TF diffusion-limited  association rate
$k_{D}$. The latter is defined as the
rate at which TFs, starting from an equilibrium distribution, arrive
at (and instantly bind) the promoter. $k_{\rm D}$ is a complicated
function of the diffusion speed of the TF in the cytoplasm and along the
DNA, the rate of non-specific TF-DNA binding, the rate of TF-DNA
dissociation and the TF-DNA binding cross-section. In terms of $k_D$ and $k_+$, the escape
probability can be written as
\begin{equation}
 S_{\mathrm{rad}}(\infty|z_0) = \frac{k_D}{k_+ + k_D} 
\elabel{Sinfty}
\end{equation}
which yields for the correlation time:
\begin{equation}
 \tau_{n} = \frac{k_+ + k_D}{(k_+ \bar{\xi} + k_-) k_D}.
\end{equation}
In \aref{noise:rates} we also show that the effective association rate
$k_{\rm on} = k_{\rm rad}(t\to \infty)$ and the effective dissociation
rate $k_{\rm off}$ are given by the diffusion-limited rate
$k_D$ and the intrinsic binding and
unbinding rates $k_+$ and $k_-$:
\begin{eqnarray}
 \frac{1}{k_{\mathrm{on}}} &=& \frac{1}{k_+} +
 \frac{1}{k_{\mathrm{D}}},\elabel{kon}\\
 \frac{1}{k_{\mathrm{off}}} &=& \frac{1}{k_-} + \frac{K_{\mathrm{eq}}}{k_{\mathrm{D}}},\elabel{koff}
\end{eqnarray}
where $K_{\rm eq} \equiv k_+/k_- = k_{\rm on} / k_{\rm
  off}$ is the equilibrium constant. The correlation time can be
expressed in terms of these rates as
\begin{equation}
  \tau_{n} = \frac{1}{k_{\mathrm{on}} \bar{\xi} + k_{\mathrm{off}}}.
\elabel{tau_n}
\end{equation}

To summarize, once we have $k_D$, we can find from the expressions
above the long-time limit of $S_{\mathrm{rad}}(t|z_0)$, the effective
association and dissociation rates $k_{\rm on}$ and $k_{\rm off}$, as
well as the correlation time $\tau_n$. In section \ref{sec:Model} we
show how we can obtain the diffusion-limited promoter association rate
$k_D$ for a TF that can diffuse in the cytoplasm, slide along the DNA,
and bind non-specifically to the DNA. The above analysis pertains,
however, also  to
other problems in which signaling molecules have to bind a receptor
molecule, possibly involving rounds of 3D, 2D or
1D diffusion; the different scenarios only yield different expressions
for the diffusion-limited arrival rate of the signaling molecules at the
receptor molecule, $k_D$.

\subsection{The sensing error}
Using the expression for the variance in our estimate of $\bar{n}$, \eref{std_devT}, in
combination with the result of \eref{CT1}, we find the
general expression for the fractional error in our estimate of the
promoter occupancy
\begin{equation}
 \left( \frac{\delta n}{\bar{n}} \right)^2 = 2 \, \frac{\sigma_n^2}{\bar{n}^2} \, \frac{\tau_n}{T} = 2 \, \frac{\sigma_n^2}{\bar{n}^2} \, \frac{\tau_{\mathrm c}}{S_{\mathrm{rad}}(\infty|z_0)} \, \frac{1}{T}.
 \elabel{EqProNoise}
\end{equation}
We combine equations \eref{Sinfty} and \eref{EqProNoise} to find a
general relation for the estimation error in terms of rate constants:
\begin{eqnarray}
 \elabel{NoisePro}
 \left( \frac{\delta n}{\bar{n}} \right)^2 & = & 2 \bar{n}(1-\bar{n})\left[\frac{1}{\bar{n} k_D T \bar{\xi}} + \frac{1-\bar{n}}{k_- T \bar{n}^2} \right] \\
 \elabel{NoisePro2}
 & = & \frac{2 \sigma_n^2}{T \bar{\xi} \bar{n}} \, \frac{1}{k_{\mathrm{on}}},
\end{eqnarray}
where we have used that $\bar{n}k_- = (1-\bar{n})k_+\bar{\xi}$. A cell
has to estimate the average TF concentration on the DNA, $\bar{\xi}$,
from the average promoter occupancy $\bar{n}$. The fluctuations in the
concentration estimate are related to the fluctuations in the
promoter-occupancy estimate via
\begin{equation}
 \delta \xi = \left|\frac{\partial \xi}{\partial n}\right| \delta n \quad \Rightarrow \quad  \delta \xi = \frac{\bar{\xi}}{\bar{n}(1-\bar{n})}\delta n,
 \elabel{EqxiNoise}
\end{equation}
and therefore the error in the concentration inferred from the
promoter state becomes
\begin{eqnarray}
 \elabel{NoiseXi}
 \left( \frac{\delta \xi}{\bar{\xi}} \right)^2 & = & \frac{2}{\bar{n}(1-\bar{n})} \left( \frac{\bar{n}}{k_D T \bar{\xi}} + \frac{1-\bar{n}}{k_- T}  \right) \\
 \elabel{NoiseXi2}
  & = & \frac{2}{T \bar{\xi} (1 - \bar{n})} \, \frac{1}{k_{\mathrm{on}}}.
\end{eqnarray}
This expression has an intuitive interpretation: the fractional error
in the concentration estimate decreases with the number of binding events during the integration
time $T$, which is given by the number of binding events if the promoter
were always free, 
$\bar{\xi}\,k_{\mathrm{on}}\,T$, times the fraction of time it is indeed free, $1-\bar{n}$. 

To derive the error in the estimate of the concentration in the
cytoplasm, we can exploit a detailed-balance relation for
the TF concentration on the DNA, $\bar{\xi}$, and that in the cytoplasm,
$\bar{c}$: $k_d \bar{\xi} = k_a \bar{c}$. Here, $k_d$ is the rate at
which a TF dissociates from the DNA to which it was bound
non-specifically, and $k_a$ is the rate at which it associates with
the DNA (non-specifically). Using this relation,  the expression for the
fractional error in the cytoplasmic concentration estimate becomes
\begin{eqnarray}
 \elabel{NoiseC} 
 \left( \frac{\delta c}{\bar{c}} \right)^2 & = & \frac{2}{\bar{n}(1-\bar{n})} \left( \frac{k_d}{k_a} \, \frac{\bar{n}}{k_D T \bar{c}} + \frac{1-\bar{n}}{k_- T}  \right) \\ \elabel{NoiseC2} 
 & = & \frac{2}{T \bar{c} (1 - \bar{n})} \, \frac{k_d}{k_a} \, \frac{1}{k_{\mathrm{on}}}.
\end{eqnarray}
Lastly, we point out that the first term on the right-hand side of
\erefsthree{NoisePro}{NoiseXi}{NoiseC} gives the contribution to the
sensing error from the finite speed of diffusion, while the second
term gives the contribution from the intrinsic promoter switching
dynamics.

\subsection{The assumptions of our theory}
\label{sec:Assumptions}
Here we discuss the assumption, \eref{Def_Ssi}, and the approximation
of our theory, \eref{Seq_approx},  in
more detail.

\eref{Def_Ssi} states that after each TF dissociation event,
the other TFs have the equilibrium distribution. By combining \eref{std_devT} and \eref{LapCorr}, it
can be seen that this assumption implies that the correlation time of
the promoter is given by
\begin{eqnarray}
\tau_n &=& \bar{n} \hat{\mathscr{S}}_{\mathrm{rad}}(s=0|{\mathrm{eq}}),\\
&=&\bar{n}\tau_{\rm off}
\end{eqnarray}
where $\tau_{\rm off} = \int_0^\infty
\mathscr{S}_{\mathrm{rad}}(t|{\mathrm{eq}}) dt=
\hat{\mathscr{S}}_{\mathrm{rad}}(s=0|{\mathrm{eq}})$ is the mean
unbound time of a free promoter surrounded by TFs obeying the
equilibrium distribution. The fact that the correlation time $\tau_n$
depends on the mean off time $\tau_{\rm off}$ and the mean occupancy
$\bar{n}=\tau_{\rm on} / (\tau_{\rm on}+\tau_{\rm off})$ (and thus the
mean on time $\tau_{\rm on}$), but not on the history of binding
events, is a direct consequence of our assumption that after each TF
dissociation event, the other TFs have the equilibrium distribution.

The mathematical approximation,
\eref{Seq_approx}, implies that $\tau_{\rm
  off}=\hat{\mathscr{S}}_{\mathrm{rad}}(s=0|{\mathrm{eq}})=1/(k_{\rm
  on}\bar{\xi})$. This is the mean waiting time for a Markov binding
process with rate $k_{\rm on}\bar{\xi}$. While approximation
\eref{Seq_approx} does not assume that binding is Markovian for all
times, it does imply that in the relevant long-time limit binding
occurs with a constant rate, yielding
$\mathscr{S}_{\mathrm{rad}}(t|{\mathrm{eq}})=e^{-k_{\rm
    on}\bar{\xi}t}$. 

Our theory predicts that the promoter correlation time $\tau_n$ is that of
a two-state Markov state model, in which the switching events are
independent, the waiting times are uncorrelated and exponentially
distributed, and the promoter switches in a memoryless fashion with
rates $k_{\rm on} \bar{\xi}$ and $k_{\rm off}$ that are constant in
time. Therefore $\tau_n = (k_{\rm on}\bar{\xi}+k_{\rm off})^{-1}$. Below, we will 
see that that in the relevant long-time limit the
promoter indeed switches in a Markovian fashion between the TF bound
and unbound state.

\subsection{Optimizing sensing precision by minimizing the search time}
We now address the question whether the system can maximize the
sensing precision by optimizing the strength of non-specific DNA
binding, characterized by the equilibrium constant $K_{\rm eq}^{\rm ns} = k_a /
k_d$. It is important to realize that the TF concentration in the cytoplasm,
$\bar{c}$, and the TF concentration on the DNA, $\bar{\xi}$, are related via
the detailed-balance relation $k_a \bar{c} = k_d \bar{\xi}$. This
means that if were to fix $\bar{c}$, raising the DNA affinity $k_a /
k_d$ would increase $\bar{\xi}$, and hence the total number of TFs in
the system. This would trivially reduce the sensing error. The interesting question is whether there
is an optimal DNA-binding strength that minimizes the sensing error
for a fixed total number of TFs, $N$.

Since the TFs are either in the cytoplasm with a volume $L^3$, or
nonspecifically bound to the DNA with a length $L_{\mathrm{D}}$, this
yields the following constraint on the number of TFs:
\begin{eqnarray}
 N &=&  \bar{c} L^3 + \bar{\xi} L_{\mathrm{D}}, \\
&=& \bar{c}(L^3 + K_{\rm eq}^{\mathrm{ns}} L_{\mathrm{D}}),
 \elabel{FixedTF0}
\end{eqnarray}
where we have used that  $\bar{\xi} = K_{\rm eq}^{\mathrm{ns}} \bar{c}$. Combining
the above expression with \eref{NoiseC} yields:
\begin{equation}
 \frac{\delta c}{\bar{c}} = \sqrt{ \frac{2}{T (1-\bar{n})}  \,
   \frac{1}{k_\mathrm{on}} \, \frac{1}{N} \, \left( L_{\mathrm{D}} +
     \frac{L^3}{K_{\rm eq}^{\mathrm ns}} \right) }.
\elabel{NoiseCFixedTF0}
\end{equation}
Because $N=\bar{c}(L^3+K_{\rm eq}^{\mathrm{ns}} L_D)$, the expression
on the right-hand side also gives the fractional error in the estimate
of the total number of transcription factors, $\delta N/N$, and total
TF concentration. 

Interestingly, \eref{NoiseCFixedTF0} shows that 
minimizing the sensor error at fixed promoter occupancy $\bar{n}$ is equivalent to minimizing the
search time $\tau_{\mathrm{s}}$, which is the average time for a
single TF to find the promoter starting from an equilibrium
distribution:
\begin{equation}
 \tau_{\mathrm{s}} = \frac{N}{\bar{\xi}\,k_{\mathrm{on}}} =
 \frac{1}{k_{\mathrm{on}}} \left( L_{\mathrm{D}} + \frac{L^3}{K_{\rm
       eq}^{\mathrm{ns}}} \right).
\end{equation}
Indeed, the fractional error in the estimate of the number of
transcription factors as a function of the search time is
\begin{equation}
\frac{\delta N}{N} = \sqrt{\frac{2\tau_{\rm s}}{N (1-\bar{n})T}}.
\elabel{dNoN}
\end{equation}
This is one of the central results of our paper.  A system with a
minimal search time, achieves a maximal rate of uncorrelated arrivals
of TFs at the promoter. It is clear from our result in
\eref{NoiseCFixedTF0}, that the sensing error and the gene expression
noise coming from promoter-state fluctuations in such a system are
minimal. The reason why minimizing the correlation time is equivalent
to minimizing the search time is precisely that the promoter correlation
time is that of a two-state Markov model, which is determined by the
effective association rate $k_{\rm on}$ and effective dissociation
rate $k_{\rm off}$, as discussed in the previous section.

\subsection{Summary}
Before we continue with our model of promoter-TF binding, we would like to remind the reader that
we have made only one assumption up to this point, which is that after
dissociation the dissociated TF is surrounded by an equilibrium 
solution of TFs (\eref{Def_Ssi}), and one approximation, namely that
the Laplace transform of ${\mathscr{S}}_{\mathrm{rad}}(t|{\rm eq})$ is
given by \eref{Seq_approx}. We have made no assumptions on the
geometry of the system yet, such that our expression for the
correlation time and sensing precision hold for any geometry. The
above theory applies to the binding of promoter-TF binding, involving
3D diffusion and 1D diffusion, but also to the binding of signaling
molecules to proteins on the membrane, involving 3D and 2D diffusion. 
 To
obtain the correlation time and sensing precision in the different
geometries, we need to find the long-time limit of the survival
probability $S_{\mathrm{rad}}(\infty|z_0)$ or the diffusion-limited
on-rate for a single particle, $k_{D}$, in these different
scenarios. Only one of these quantities suffices as both are related
via \eref{Sinfty}.  Deriving $S_{\mathrm{rad}}(\infty|z_0)$ and $k_D$
for promoter-TF binding will
be our main goal of the next section.

\subsection{Model}
\label{sec:Model}
We now derive the long-time limit of $S_{\mathrm{rad}}(t|z_0)$,
$S_{\mathrm{rad}}(\infty|z_0)$, for the model shown in
\fref{promoter_schematic}. The DNA near the promoter is described as a
straight cylinder.  In the cytoplasm TFs diffuse with diffusion
constant $D_3$. A TF molecule can (non-specifically) bind DNA with an
intrinsic association rate $k_a$ when it is in contact with it; the
TF-DNA binding cross-section is $\sigma$. On the DNA, TFs can slide
with diffusion constant $D_1$, dissociate into the cytoplasm with the
intrinsic dissociation rate $k_d$, or, when they arrive at the
promoter, bind the promoter with the intrinsic association rate
$k_+$. A promoter-bound TF can dissociate from the promoter with rate
$k_-$. We note that this model is identical to that Tka\v{c}ik and
Bialek \cite{Tkacik2009}.  From $S_{\mathrm{rad}}(\infty|z_0)$, we can
obtain $k_D$, $k_{\rm on}$, $k_{\rm off}$,  $\tau_n$, and the sensing error via \erefs{Sinfty} -
\ref{eqn:tau_n} and \eref{NoiseC}.

\begin{figure}
\centering
\includegraphics[scale=.45]{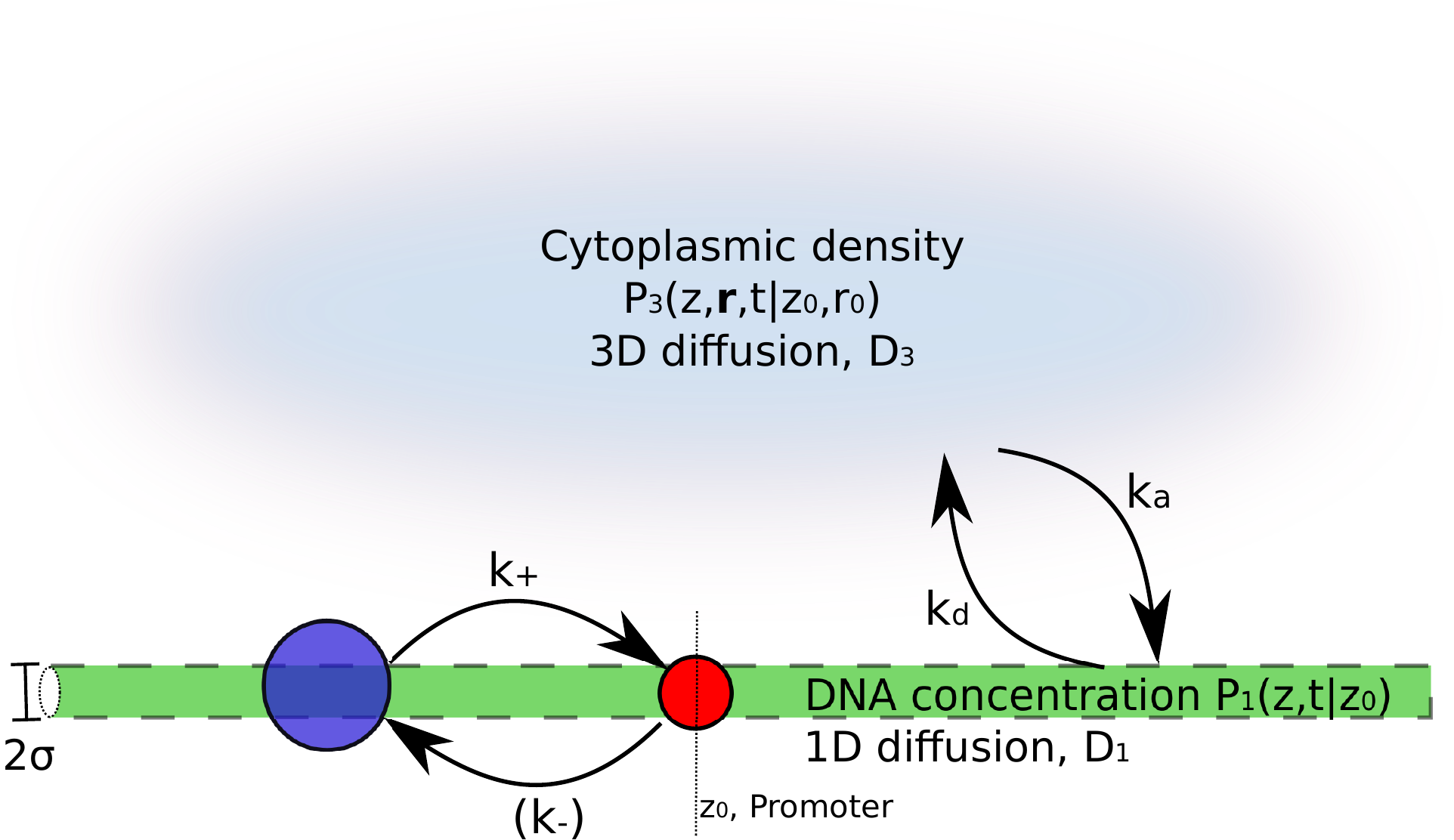}
\caption{\flabel{promoter_schematic} Model of TFs that can bind the
  promoter via 3D diffusion in the cytoplasm and 1D diffusion along
  the DNA. The DNA near the promoter is modeled as a straight
  cylinder. In the cytoplasm, the TFs diffuse with diffusion constant
  $D_3$. A single TF can associate with the DNA with the intrinsic
  association rate $k_a$ when it is in contact with it. On the DNA, a
  TF can slide with diffusion constant $D_1$, dissociate into the
  cytoplasm with the intrinsic dissociation rate $k_d$, or, when it
  arrives at the promoter, bind the promoter with rate $k_+$. A
  promoter-bound TF can dissociate from it with rate $k_-$.  The
  diffusion along the DNA is described with the Green's function
  $P_1(z,t|z_0)$, and the diffusion inside the cytoplasm with
  $P_3(z,{\mathbf r},t|z_0,r_0)$. To derive $S_{\mathrm{rad}}(t|z_0)$,
  we consider a single TF that starts at contact with the promoter,
  denoted by $z_0$.  $S_{\mathrm{rad}}(t\to \infty|z_0)$, the
  diffusion-limited binding rate $k_D$, the promoter correlation time
  $\tau_n$, and the sensing precision can be obtained
  via  \erefs{Sinfty} -
\ref{eqn:tau_n} and \eref{NoiseC}.}
\end{figure}

To calculate $S_{\mathrm{rad}}(\infty|z_0)$, we write down the full set of
diffusion equations governing the behavior of a single TF starting on
the promoter site:
\setlength{\jot}{12pt}
\begin{eqnarray}
\elabel{DE_full}
\pdiff{P_{1}(z,t|z_0)}{t} & = & D_1 \frac{\partial^2 P_{1}(z,t|z_0)}{\partial^2 z} - k_d P_{1}(z,t|z_0) \nonumber \\ 
- k_+ P_{1}(z,t|z_0) \delta(z - z_0) & + & k_a P_{3}(z, |{\mathbf r}| = \sigma, t|z_0, r_0) \\
\pdiff{P_{3}(z,{\mathbf r},t|z_0,r_0)}{t} & = & D_3 \nabla^2 P_{3}(z, {\mathbf r}, t|z_0,r_0) \nonumber \\ 
- \Big[ k_a P_{3}(z, {\mathbf r}, t|z_0,r_0) & - & k_d P_{1}(z,t|z_0) \Big] \frac{\delta(|{\mathbf r}| - \sigma)}{2 \pi \sigma}.
\end{eqnarray}
Here $P_1(z, t|z_0)$ is the Green's function describing the 1D sliding
of the TF along the DNA, starting at the promoter positioned at
$z_0$.  Excursions in
the cytoplasm are described by $P_3(z, {\mathbf r}, t|z_0,r_0)$, where
$r_0=0$, stating that the particle starts on the DNA. We model the DNA
as an infinitely long rod along the z-axis. Because the TF-DNA
cross-section is $\sigma$, the probability exchange between the DNA and bulk
happens at a distance $\sigma$ from the z-axis, imposed by the delta
function in the second equation. In order to solve the equations, we
first Laplace transform them with respect to time
\begin{eqnarray}
 s \, \hat{P}_{1}(z|z_0) - \delta(z - z_0) & = & D_1 \frac{\partial^2\hat{P}_1(z|z_0)}{\partial^2 z}  - k_+ \hat{P}_1(z|z_0) \delta(z - z_0) \nonumber
\\ & - & k_d \hat{P}_1(z|z_0) + k_a \hat{P}_3(z, |{\mathbf r}| = \sigma|z_0,r_0) \nonumber \\
 s \, \hat{P}_3(z,{\mathbf r}|z_0,r_0) & = & D_3 \nabla^2 \hat{P}_3(z, {\mathbf r}|z_0,r_0) \nonumber \\
 & - & \left[ k_a \hat{P}_3(z,{\mathbf r}|z_0,r_0) - k_d \hat{P}_1(z|z_0) \right] \frac{\delta(|{\mathbf r}| - \sigma)}{2 \pi \sigma}, \nonumber
\end{eqnarray}
where we explicitly included the initial condition of one particle placed in contact with the promoter 
site on the DNA by the Dirac delta function.  We continue by Fourier transforming with respect to space
\begin{eqnarray}
 s \, \tilde{P}_1(q|z_0) - 1 &=& - D_1 q^2 \tilde{P}_1(q|z_0) - k_+ \hat{P}_1(z_0|z_0) \\ 
&& - k_d \tilde{P}_1(q|z_0) + k_a \tilde{P}_3(q|z_0,r_0) \elabel{FDE_dna} \nonumber \\
 s \, \tilde{P}_3(q,{\mathbf k}|z_0,r_0) &=& - D_3 (q^2 + k^2) \tilde{P}_3(q, {\mathbf k}|z_0,r_0)  \\ 
&&- \left[ k_a \tilde{P}_3(q|z_0,r_0) - k_d \tilde{P}_1(q|z_0) \right] J_{0}(k \sigma). \nonumber  
 \elabel{FDE_bulk}
\end{eqnarray}
Here $q$ is the spatial Fourier variable conjugate to $z$, and
${\mathbf k}$ is conjugate to ${\mathbf r}$. $J_0(k\sigma)$ is the
zeroth order Bessel function of the first kind. We take both the
promoter and initial position to be at the origin: $z_0 = 0$. We want
to solve these equations for $\tilde{P}_1(q|z_0)$, from which we can
extract the required survival probability
$S_{\mathrm{rad}}(\infty|z_0)$. Observe that the cytoplasmic density
$\tilde{P}_3$ in \eref{FDE_dna} is a function of $q$ only (and not of
${\mathbf k}$). In order to
solve for $\tilde{P}_1$, we need an expression for
$\tilde{P}_3(q|z_0,r_0)$ in terms of $\tilde{P}_1$. We start by
solving the second equation for $\tilde{P}_3(q,{\mathbf k}|z_0,r_0)$,
\begin{equation}
 \tilde{P}_3(q,{\mathbf k}|z_0,r_0) = \frac{k_d \tilde{P}_1(q|z_0) - k_a\tilde{P}_3(q|z_0,r_0)}{s + D_3(q^2 + k^2)} J_{0}(k \sigma ).
\end{equation}
Fourier back-transforming both sides of the equation in ${\mathbf k}$, at $r=\sigma$, 
where we implicitly integrate over all ${\mathbf r}$,
\begin{eqnarray}
 \elabel{FourierBackC}
 \tilde{P}_3(q|z_0,r_0) &=& \int d\nu \frac{\delta(r-\sigma)}{2\pi r} \tilde{P}_3(q,{\mathbf k}|z_0,r_0) \\
 & =& \frac{k_d \tilde{P}_1(q|z_0) - k_a\tilde{P}_3(q|z_0,r_0)}{2 \pi D_3} \, I_{0}\left( \chi \right) K_{0}\left( \chi \right) \nonumber,
\end{eqnarray}
where $d\nu = \frac{d^2 k}{(2 \pi)^2} \mathrm{e}^{-i\,{\mathbf k}\cdot{\mathbf r}}$, 
$I_0$ and $K_0$ are the zeroth order modified Bessel functions of the first and 
second kind respectively, and $\chi=\sigma \sqrt{q^2 + \frac{s}{D_3}}$. Solving the above for 
$\tilde{P}_3(q|z_0,r_0)$, and substituting the result into \eref{FDE_dna}, 
we obtain the solution for $\tilde{P}_1(q|z_0)$. Again, back-transforming this equation in $q$ at 
the position of the promoter, $z_0=0$, we find
\begin{equation}
 \hat{P}_1(z_0,s|z_0) = \int \frac{d\,q}{2 \pi} \frac{1 - k_+ \hat{P}_1(z_0,s|z_0)}{s + D_1 q^2+k_d F^{-1}(q,s)}
 \elabel{FDE_dna2}
\end{equation}
where
\begin{equation}
F(q,s) = 1 + \frac{k_a}{2 \pi D_3} \, I_{0}\left( \chi \right) K_{0}\left( \chi \right).
\end{equation}

Finally, we can solve \eref{FDE_dna2} for $\hat{P}_1(z_0, s|z_0)$ to obtain the probability density at the promoter site in Laplace space. In the limit $s \to 0$, our expression becomes
\begin{equation}
 \lim_{s \to 0} \hat{P}_1(z_0, s|z_0) = \frac{I(\alpha,\beta)}{\pi D_1 / \sigma + k_+ \, I(\alpha,\beta)}
\end{equation}
where
\begin{eqnarray}
 I(\alpha,\beta) &=& \int_0^{\infty} \frac{dt}{t^2 + \beta[1 + \alpha \, I_0(t) K_0(t)]^{-1}} \elabel{Int} \\
 \alpha &=& \frac{k_a}{2 \pi D_3} \elabel{EqAlpha} \\
 \beta &=& \frac{\sigma^2 k_d}{D_1} \elabel{EqBeta}.
\end{eqnarray}
To relate this result to the large time limit of the survival
probability $S_{\mathrm{rad}}(\infty|z_0)$, we exploit that the flux
into the promoter at any given time is $k_+\,P_1(z_0,t|z_0)$, and that
the total flux which leaks away through the promoter is equal to the
integral over all times of the flux. Since the $s\to0$ limit in the
Laplace transformed function $\hat{P}_1(z_0,s|z_0)$ is exactly this
integral, we find the survival probability via
\begin{eqnarray}
  \lim_{t \to \infty}S_{\mathrm{rad}}(t|z_0) &=& 1 - k_+ \, \int_0^{\infty} P_1(z_0, t|z_0) dt \nonumber \\
 &=& 1 - k_+ \lim_{s \to 0} \hat{P}_1(z_0, s|z_0) \nonumber \\ 
 &=& \frac{\frac{\pi D_1}{\sigma I(\alpha,\beta)}}{\frac{\pi D_1}{\sigma I(\alpha,\beta)} + k_+}.
\end{eqnarray}
Comparing with \eref{Sinfty}, the diffusion limited rate constant is
\begin{equation}
 k_D = \frac{\pi D_1}{\sigma I(\alpha,\beta)}.
 \elabel{kDFullJ}
\end{equation}
Plugging this result into \erefstwo{NoisePro}{NoiseC}, the fractional
error in the promoter-occupancy estimate is
\begin{equation}
 \elabel{NoiseProFull}
 \left( \frac{\delta n}{\bar{n}} \right)^2 = 2 \bar{n}(1-\bar{n})\left[\left( \frac{\sigma I(\alpha,\beta)}{\bar{n} \pi D_1 T \bar{\xi}} \right)+\frac{1-\bar{n}}{k_- T \bar{n}^2} \right],
\end{equation}
and that in the cytoplasmic concentration $\bar{c}$ is
\begin{equation}
 \elabel{NoiseCFull} 
 \left( \frac{\delta c}{\bar{c}} \right)^2 = \frac{2}{\bar{n}(1-\bar{n})} \left( \frac{k_d}{k_a} \, \frac{\bar{n} \sigma I(\alpha,\beta)}{\pi D_1 T \bar{c}} + \frac{1-\bar{n}}{k_- T}  \right).
\end{equation}

In the limit that DNA binding is reaction limited, $k_a\ll D_3$, it is very unlikely that the TF will rebind with the DNA after falling off, and the cytoplasm becomes effectively well mixed. In this limit, $\alpha\to0$ in the integral of \eref{Int}, and we can analytically solve it. The diffusion limited on-rate to the promoter becomes in this limit
\begin{equation}
k_D = \sqrt{4 D_1 k_d} = \sqrt{2} b k_d,
\elabel{kD_WM}
\end{equation}
where $b=\sqrt{2 D_1/k_d}$ is the average length of a single excursion
along the DNA. This equation has an intuitive interpretation. On
average, a TF binding the DNA within a distance $\sim b$ from the
promoter site, will find it. The rate
at which molecules leave the DNA from this region is $~\sim b k_d
\bar{\xi}$.  Because our system obeys detailed balance, this rate of
departure equals the rate of arrival, $k_D \bar{\xi}$, hence $k_D \sim b
k_d$.

\section{RESULTS}

\subsection{Comparing theory with simulations}
To test our theory, we have performed simulations 
using the enhanced Green's Function Reaction Dynamics algorithm
(GFRD) \cite{Takahashi2010}. Recently, we have expanded the functionality
of GFRD\, to simulate diffusion and reactions on a plane (2D) and
along a cylinder (1D). Particles can exchange between the bulk
and planes or cylinders via association and dissociation. Furthermore, 
specific binding sites can be added to a cylinder to which
a particle diffusing along the cylinder can bind. Importantly, GFRD,
is an exact scheme for simulating reaction-diffusion problems at the
particle level, making it ideal to test theoretical predictions.

\begin{center}
    \begin{table}[t]
	\begin{tabular}{  l  l  l }
	\hline 
	\textbf{Parameter} & \textbf{Value} & \textbf{Motivation} \\ \hline
	$L$ & 1 $\upmu {\rm m}$ & Bacterium size \\ 
	$L_{\mathrm{D}}$ & 1 mm & E. Coli DNA length \\ 
	$\mathrm{TF}_0$ & 10 & \cite{Elf2007} \\ 
	$D_1$ & 5 $10^{-2}$ $\upmu {\rm m^2/s}$ & \cite{Elf2007} \\ 
	$D_3$ & 3 $\upmu {\rm m^2/s}$ & \cite{Elf2007} \\ 
	$k_a$ & 1 $\upmu {\rm m^2/s}$ & \cite{Tabaka2014} \\ 
	$k_d$ & 1000/s & \cite{Elf2007,Tabaka2014} \\ 
	$k_+$ & Varies & - \\ 
	$k_-$ & 100/s in \fref{PowerPlot} and \fref{NoisePlotFull} & - \\ 
	\phantom{$k_-$} & Varies in \fref{PromoterNoise_kdD} and \fref{DNAOcc_kaP} & Such that $\bar{n}=0.5$ \\ 
	$\sigma$ & 4 nm & - \\ 
	$T$ & 100 s & - \\ 
	\hline 
	\end{tabular}    
    \caption{\tlabel{parameters} Typical values of the parameters used
      in our simulations and figures. When different values for the
      parameters are used, they are given in the text or 
      figure captions.}
    \end{table}
\end{center}

Our simulation setup consist of a box with periodic boundary
conditions. To model the DNA, the box contains a cylinder, which
crosses the box. The promoter is modeled as a specific binding site at the middle of the
cylinder. The box contains 10 transcription factors. Other details, such as
parameter values, are given in \tref{parameters}. We record the
trajectory of the promoter, switching between the occupied and
unoccupied state, for a period of 3000 seconds.

The key quantity of our theory is the zero-frequency limit of the
power spectrum, $\lim_{\omega\to 0}P_n(\omega)=2\sigma^2_n \tau_n$,
since the uncertainty in the promoter-occupancy and the concentration
estimate can be directly obtained from this quantity and the gain (see
\eref{std_devT} and \eref{gain}). We therefore take the power spectrum
of the promoter signal, following the procedure described in
\cite{VanZon2006}.

\begin{figure}
\includegraphics[scale=1.0]{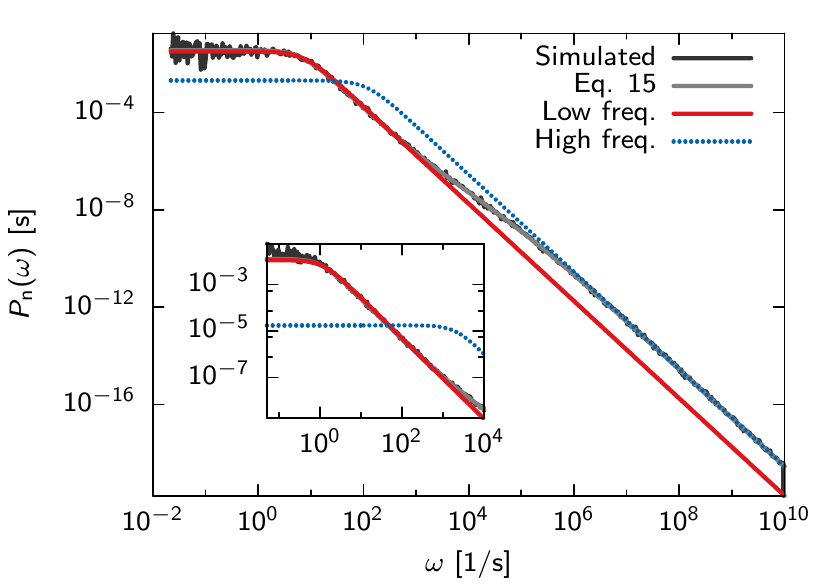}
\caption{\flabel{PowerPlot} The power spectrum of the 
  promoter state $P_n(\omega)$, for $\bar{n}=0.14$. The simulation results (\textit{black
    line}) agree well with the theoretical prediction 
  (\eref{Corr_approx}). At high frequencies, the effect of
  diffusion is negligible and the promoter dynamics is that a
  Markovian switching process with intrinsic rates $k_+\bar{\xi}$ and
  $k_-$ (blue dotted line). At low frequencies, promoter switching can
  again be described by a random telegraph process, but now with
  effective rates $k_{\rm on}\bar{\xi}$ and $k_{\rm off}$ (red solid line). 
The association rate $k_+ =
  0.16\mathrm{mm/s}$ and other parameters as given in Table
  \tref{parameters}. The inset shows a
  power spectrum for a higher association rate $k_+=19  
  \mathrm{mm/s}$ where $\bar{n}=0.67$.}
\end{figure}

\fref{PowerPlot} shows that the agreement between theory and
simulations is very good over essentially the full frequency range, as
observed previously for the binding of ligand to a spherical receptor
\cite{Kaizu2014}.  In the high-frequency regime, diffusion hardly
plays any role and the receptor dynamics is dominated by the binding
of TF molecules that are essentially in contact with the promoter;
consequently, the power spectrum is well approximated by that of a
binary switching process with uncorrelated exponentially distributed
waiting times with the intrinsic correlation time
$\tau_c=(k_+\bar{\xi}+k_-)^{-1}$ (blue dotted line). The theory
also accurately describes the intermediate frequency regime, in which
a dissociated TF molecule manages to
diffuse away from the promoter, but then rebinds it before another TF
molecule does.    The low frequency regime of the power spectrum corresponds to
the regime in which after promoter dissociation the TF molecule
diffuses into the bulk and, most likely, another TF molecule from the
bulk binds the promoter. In this regime, the spectrum is well
approximated by that of a memoryless switching process with the same
effective correlation time as that of our theory,
$\tau_n=(k_{\rm on}\bar{\xi}+k_{\rm off})^{-1}$
(solid red line).

\fref{NoisePlotFull} shows the zero-frequency limit of the power
spectrum, $P_n(\omega\to 0)$, as a function of the
the average occupancy $\bar{n}$, where we change $\bar{n}$ by varying
the intrinsic association rate $k_+$.  The theory matches simulation very well up to
$\bar{n}\sim0.8$. For higher value's of $\bar{n}$, it is harder to
measure the plateau value at the low frequency limit of the power
spectrum, as shown in the inset of \fref{PowerPlot}.

\begin{figure}
\centering
\includegraphics[scale=1.0]{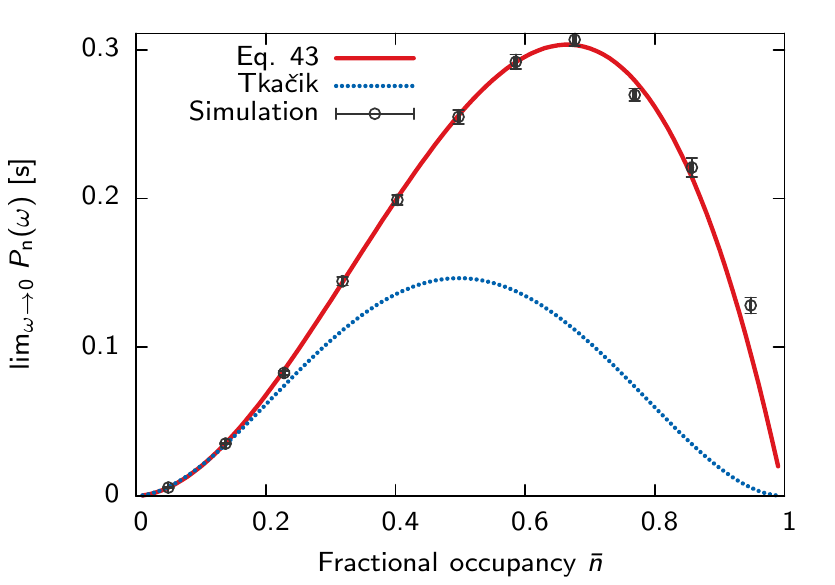}
\caption{\flabel{NoisePlotFull} The low frequency limit of the power
  spectrum given by \eref{std_devT} as a function of the average
  occupancy $\bar{n}$ (red line), is in excellent agreement with
  simulation results.  The dashed (blue) line shows
  the low frequency limit as derived in \cite{Tkacik2009}, which is
  symmetric in $\bar{n}$. $\bar{n}$ is varied by varying $k_+$. 
  Other parameters are given in \tref{parameters}.}
\end{figure}

\subsection{Why the theory is accurate: timescale separation}
\label{sec:ValidityAssumptions}
The key assumption of our theory is \eref{Def_Ssi}, which states after
each TF-promoter dissociation event, the other TFs have the
equilibrium distribution. This assumption breaks down when two
conditions are met: a) the rebinding of a TF to the promoter is
pre-empted by the binding of a second TF from the bulk; {\em and} b)
when the second TF dissociates from the promoter before the first has
diffused in the bulk \cite{Kaizu2014}. We now consider both
conditions.

In {\em E. coli}, the time required for a lac repressor molecule to
bind the promoter from the bulk is on the other of seconds to minutes
\cite{Hammar2012a}. The time a dissociated repressor molecule spends
near the promoter is on the order of the sliding time, which is $1-100
{\rm ms}$ \cite{Elf2007,Marklund2013}. This timescale separation means
that the likelihood that a TF from the bulk pre-empts the rebinding of
a dissociated TF to the promoter is negligible; the probability of
rebinding interference is very low and a dissociated TF rebinds the
promoter before escaping into the bulk as often as when it would be
the only TF in the system. This means that condition a) is not
satisfied, and hence the assumption of \eref{Def_Ssi} holds.

Even if there is occasionally rebinding interference, then
\eref{Def_Ssi} still probably holds because condition b) is not
met. To determine whether a TF dissociates from the promoter before
the previously dissociated TF has escaped into the bulk, we compare
$k_-^{-1}$, the time a TF is specifically bound to the promoter, to
the time a TF resides on the DNA (bound non-specifically) before
  escaping into the bulk.  We can estimate the intrinsic
dissociation time $k_-$ from the specific dissociation constant
$K_D^{s}=k_d k_-/(k_+k_a)$ and from $k_d$, $k_+$ and $k_a$. The
microscopic non-specific binding rate for the lac repressor has been
estimated to be $k_i \approx 3 \times 10^5 {\rm M^{-1}s^{-1}}$
\cite{Tabaka2014}. This yields $k_a = k_i/d \approx 1 \mu {\rm
  m^2/s}$, where $d=0.3{\rm nm}$ is the distance between DNA base
pairs.  The specific promoter association rate $k_+$ can be estimated
from $k_+=\gamma D_1 / d$, which with $\gamma = 0.1$
\cite{Hammar2012a} and the 1D diffusion constant $D_1=0.05 \mu{\rm
  m}^2/{\rm s}$ \cite{Elf2007}, yields $k_+\approx 10 \mu{\rm
  m/s}$. The DNA dissociation rate for the lac repressor is $k_d
\approx 10-1000 /{\rm s}$ \cite{Elf2007,Marklund2013}. The
dissociation constant for repressor binding to the operator $O_1$ is
in the nM regime \cite{Riggs1970}. Taken together, these numbers imply
that the time the repressor is bound for a time $k_-^{-1}$ that is at
least seconds. This is consistent with the experimental observation of
Hammar et al. that individual operator-bound LacI molecules appear as
diffraction-limited spots on a $4 {\rm s}$ timescale
\cite{Hammar2012a}. This is longer than our estimate for how long a TF
which has dissociated from the promoter, resides near the promoter
before escaping into the bulk, which is $1-100 {\rm ms}$
\cite{Elf2007,Marklund2013}. We thus conclude that also condition b)
is not satisfied; even if rebindings occur and condition a) is met,
the central assumption of our theory, \eref{Def_Ssi}, will thus hold.

The principal reason why the key assumption \eref{Def_Ssi} holds, is
thus that the time TFs spend near the promoter is very short, both on
the timescale at which TFs arrive at the promoter from the bulk and on the
timescale a TF is bound to the promoter. 

That TFs spend little time near the promoter as compared to the time
required to bind the promoter from the bulk, is also the reason why
the mathematical approximation, \eref{Seq_approx}, is very accurate. In this
approximation,
$\mathscr{S}_{\mathrm{rad}}(t|{\mathrm{eq}})=e^{-\bar{\xi}k_{\rm
    on}t}$ at long times. The range over which
$\mathscr{S}_{\mathrm{rad}}(t|{\mathrm{eq}})=e^{-\bar{\xi} \int_0^t
  dt^\prime k_{\rm rad}(t^\prime)}$ deviates from this long-time limit
  is determined by how rapidly $S_{\rm rad}(t|z_0)$ decays
(because that determines how fast $k_{\rm rad}(t)$ reaches its
long-time limit $k_{\rm on}$, see \eref{krad}). This decay is
dominated by $k_d$, which is at least an order of magnitude faster
than the long-time decay governed by $k_{\rm on}\bar{\xi}$.  Hence,
after a promoter dissociation event, the dissociated TF essentially
instantly rebinds the promoter or instantly escapes into the bulk, and
then (most likely) another TF binds the promoter in a memoryless
fashion, with a constant rate $k_{\rm on}\bar{\xi}$.

\subsection{Comparison with Tka$\check{\mathbf c}$ik and Bialek}
The sensing precision was derived earlier by Tka$\check{\mathrm c}$ik
and Bialek, but via a different method \cite{Tkacik2009}. They start with the
differential equations governing the fluctuations in the promoter
state $\delta n$, and relate these to changes in the
free energy due to the binding and unbinding of
TFs. The fluctuations in the occupancy are then related to the power
spectrum via the fluctuation-dissipation theorem.

Their final result for the noise in the promoter state is (Eq. 68 in \cite{Tkacik2009}) 
\begin{equation}
 \elabel{NoiseProTka}
 \left( \frac{\delta n}{\bar{n}} \right)^2 = 2 \bar{n}(1-\bar{n})\left[ (1 -\bar{n}) \, \frac{\sigma I(\alpha,\beta)}{\bar{n} \pi^2 D_1 T \bar{\xi}} + \frac{1-\bar{n}}{k_- T \bar{n}^2} \right]
\end{equation}
where $I(\alpha,\beta)$ is given by
\begin{equation}
 \elabel{IntProTka}
 I(\alpha,\beta) = \int_0^{\infty} \frac{dt}{t^2 + \beta[1 + \alpha \, {\mathrm{log}}(1 + t^{-2})]^{-1}},
\end{equation}
and $\alpha=k_a/(4 \pi D_3)$ and $\beta=k_d \sigma^2/(\pi D_1)$. 

The most important difference is the extra factor $(1-\bar{n})$ in the
diffusion term, which makes $P_n(\omega\to 0)$ 
symmetric around $\bar{n}=0.5$, as is shown in \fref{NoisePlotFull} (blue dotted line). Our simulations results in
\fref{NoisePlotFull} show however, that the maximum is reached when
the promoter is occupied for more than half of the time. 

Furthermore, in contrast to our result in \eref{NoiseCFull}, the
precision of the TF concentration estimate  (Eq. 71 in \cite{Tkacik2009})
is independent of the promoter occupancy $\bar{n}$. However, since
incoming TFs can not bind with the promoter when it is already
occupied, it becomes harder to perceive the TF concentration as the
promoter occupancy increases. In other words, the number of
independent `measurements' the promoter can make of the TF
concentration during its integration time $T$, decreases with
increasing occupancy. As a result, one would expect the noise to
diverge as $\bar{n}\to1$. Kaizu et al. \cite{Kaizu2014} obtained
precisely the same discrepancy for a spherical receptor. The extra
$(1-\bar{n})$ factor in \eref{NoiseProTka} is most
likely the result of a linearizion  \cite{Tkacik2009,Kaizu2014}.

\subsection{A coarse-grained model}
In previous work, we have shown that the effect of TF diffusion in a
spatially resolved model of promoter-TF binding can be captured in a
well-stirred model by renormalizing the association and dissociation
rates \cite{VanZon2006,Kaizu2014}. The principal observation is
that  a TF molecule near
the promoter either rapidly binds the promoter or rapidly escapes into
the bulk,  as discussed in section \ref{sec:ValidityAssumptions}. 
As a consequence, the probability that the binding of this
molecule to the promoter is pre-empted by the binding of another
ligand molecule, is negligible: a TF molecule near the promoter
binds the promoter or escapes into the bulk with splitting
probabilities that are the same as when it would be the only TF
molecule in the system. There is no (re)binding interference. This
makes it possible to integrate out the rapid rebindings and the
unsuccessful bulk arrivals, and reduce the complicated many-body
reaction-diffusion problem to a pair problem in which ligand molecules
interact with the receptor in a memory-less fashion, with renormalized
association and dissociation rates. However, in these previous
studies, the receptor (the promoter) was modeled as a sphere. While in
\cite{VanZon2006} we predicted that rebindings could also be
integrated out in a more detailed model of gene expression in which
TFs do not only diffuse in the cytoplasm but also slide along the DNA, this
question has so far not been answered. Here, we show that the answer
is positive.

When the probability of rebinding interference is negligible, the
effective dissociation rate $k_{\mathrm{off}}$  is given by \cite{VanZon2006,Kaizu2014}:
\begin{equation} k_{\mathrm{off}} = \frac{k_-}{1 + N_{\mathrm{reb}}}.
 \elabel{OffRenRates}
\end{equation}
Here $N_{\mathrm{reb}}$ is the average number of rebindings, which is
defined as the average number of rounds of rebinding and dissociation
before a dissociated TF escapes into the bulk. It is given by 
\begin{equation}
 N_{\mathrm{reb}} = \sum_{n=1}^{\infty} n \left( p_{\mathrm{reb}} \right)^n p_{\mathrm{esc}} = \frac{1-p_{\mathrm{esc}}}{p_{\mathrm{esc}}},
\end{equation}
where $p_{\mathrm{reb}}$ and
$p_{\mathrm{esc}}=1-p_{\mathrm{reb}}$ are the splitting
probabilities of a TF at contact for either rebinding the promoter or
escaping into the bulk.  The probability of a TF escaping is given by the $t \to
\infty$ limit of the survival probability of a particle starting at
contact
\begin{equation}
 p_{\mathrm{esc}} = \lim_{t \to \infty} S_{\mathrm{rad}}(t|z_0) = S_{\mathrm{rad}}(\infty|z_0).
\end{equation}
Combining the above expressions, we find that $k_{\rm off}$ is
precisely the effective dissociation rate of our theory, \eref{koff}.

When the probability of rebinding interference is negligible, the
effective association rate $k_{\rm on}$ is the rate at which a TF
arrives from the bulk at the promoter, $k_D$, times the probability
$1-S_{\mathrm{rad}}(\infty|z_0)=k_+/(k_++k_D)=p_{\rm reb}$ (see
\eref{Sinfty}) that it
subsequently binds \cite{Kaizu2014}:
\begin{equation}
k_{\rm on}=\frac{k_+k_D}{k_++k_D}.
 \elabel{OnRenRates}
\end{equation}
 This indeed is the effective association rate
of our theory, \eref{kon}. Again we see that the complicated dynamics
of 3D diffusion, 1D sliding, and exchange between cytoplasm and DNA,
is contained in the arrival rate $k_D$ and the escape probability
$S_{\mathrm{rad}}(\infty|z_0)$, which are related via \eref{Sinfty}.

This picture yields the simple two-state model:
\begin{equation}
\frac{dn(t)}{dt}=k_{\rm on}\bar{\xi}(1-n(t))-k_{\rm off}n(t).
\end{equation}
In this model, the promoter switches with exponentially distributed
waiting times between the on and the off state, with a correlation
time which is precisely that of our theory, \eref{tau_n}. As
\fref{NoisePlotFull} shows, even in the presence of 1D diffusion along
the DNA, this two-state model accurately describes the zero-frequency
limit of the power spectrum, which determines the promoter correlation
time and hence the sensing precision. The main reason why sliding does
not change our earlier result obtained for a spherical promoter
\cite{VanZon2006} is that the non-specific residence time on the DNA, $<50{\rm
  ms}$, \cite{Elf2007,Marklund2013}, is small compared to the
timescale of seconds to minutes on which TFs bind the promoter from
the bulk \cite{Hammar2012a}, see section \ref{sec:ValidityAssumptions}.

\subsection{Optimizing the sensing precision}
We now minimize the sensing error keeping the average promoter
occupancy constant at $\bar{n}=0.5$. The volume of the box is approximately that of a
bacterium such that, $L=1\mathrm{\mu m}$, and for the length of the
DNA we take the typical value $L_{\mathrm{D}}=1 \mathrm{mm}$.

In \fref{PromoterNoise_kdD} we plot the sensing error as a function of
the DNA dissociation rate $k_d$. Different lines correspond to
different values of the intrinsic promoter association rate $k_+$. In
these calculations, we fix $D_3$, $D_1$, and $k_a$, and adjust $k_-$
such that $\bar{n}=0.5$. It is seen that there is an optimal
dissociation rate $k_d$ and hence an optimal affinity $K_{\rm eq}^{\rm ns}=k_a / k_d$ 
that minimizes the sensing error. Tka\v{c}ik and 
Bialek did not find an optimum, because they did not 
constrain the total number of TFs to be constant \cite{Tkacik2009}.

Finding the promoter involves rounds of 1D diffusion along the DNA and
3D diffusion in the cytoplasm \cite{Hippel1989}. The optimal search time
is due to a trade-off between how long each round takes and the number
of rounds $M$ needed to find the promoter \cite{Hippel1989,Slutsky2004a,Hu2006}. The total search time is
$\tau_s = M (\tau_{1D} + \tau_{3D}$), where $\tau_{3D}$ is the time a
TF spends in the cytoplasm during one round and $\tau_{1D}$ is the
time it spends on the DNA \cite{Slutsky2004a,Hu2006}.  The latter is given by
$\tau_{1D}=1/k_d$. Ignoring correlations
between the point of DNA dissociation and subsequent DNA association,
$M\sim L_D / b$, where $b=\sqrt{2 D_1/k_d}$ is the average length of a
single excursion along the DNA. Hence, as $k_d$ is increased,
$\tau_{1D}$ decreases as $1/k_d$, while $M$ increases as
$\sqrt{k_d}$. This interplay leads to a minimum in the search time and
hence the sensing error.

\begin{figure}
\centering
\includegraphics[scale=1.0]{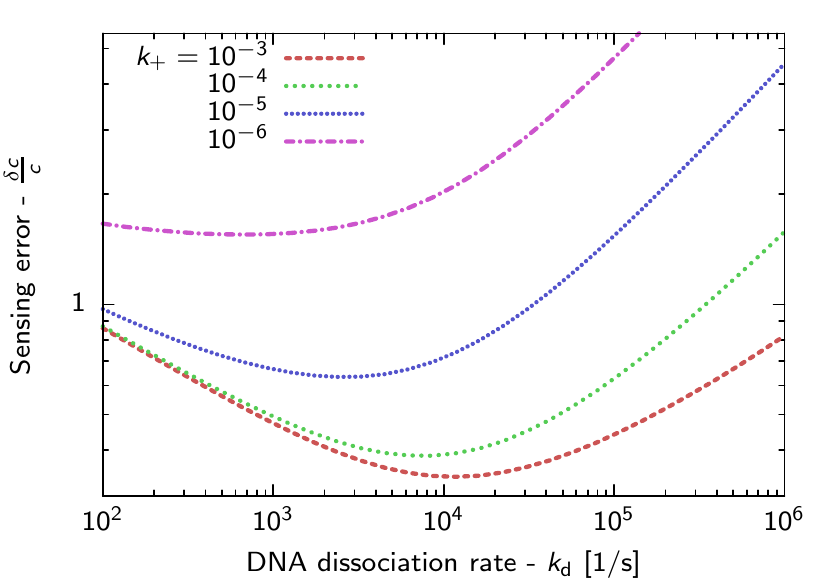}
\caption{\flabel{PromoterNoise_kdD} The sensing error
  (\eref{NoiseCFixedTF0}) as a function of the DNA dissociation rate
  $k_d$. The lines show the sensing error for different values of the
  intrinsic association rate of the promoter $k_+$, which are given in
  the legend and have units of m/s. The noise has an optimum when the
  search time of a TF for its target is the lowest. As we lower $k_+$,
  the sensing error increases because the TF has to arrive multiple times at
  the promoter before binding to it, which decreases the effective
  on-rate.  We set the integration time $T$, which is usually the
  regulated protein lifetime, to $T=100$ s. When $k_d$ is varied,
  $k_-$ is tuned such that $\bar{n}=0.5$. Other parameters are fixed
  at typical biological values, given in \tref{parameters}. }
\end{figure}

Slutsky and Mirny \cite{Slutsky2004a} predicted that for an optimal
search time, the TFs spend 50\% of their time nonspecifically bound to
the DNA and 50\% in the cytoplasm. In their model, they assumed the
cytoplasm to be well mixed ($D_3\to\infty$ in our model) and that the
search process is diffusion limited, $k_{\mathrm{on}}\to
k_D$. Recent experiments \cite{Hammar2012a}, however, have shown
that for some TFs the search process is not diffusion limited and that
therefore the intrinsic association rate to the promoter $k_+$ will be
similar or smaller than $k_D$. Furthermore, they show that a TF spends
around 90\% of its time nonspecifically bound to the DNA, much larger
than predicted. \fref{DNAOcc_kaP} supports the proposition of
\cite{Hammar2012a} that these observations are
 related.

 \fref{DNAOcc_kaP} shows the fraction of time a TF spends on the DNA,
 as a function of the association rate to the promoter $k_+$. It is 
 reasonable to assume that transcriptional regulation operates in a
 parameter regime where the sensing error is low. Therefore, for each
 point in the figure, we chose $k_d$ such that the search time is
 minimal (minimum in \fref{PromoterNoise_kdD}). At high values of
 $k_+$, promoter binding becomes diffusion limited and thus
 independent of $k_+$. For lower values of $k_+$, however, the rate of
 promoter binding becomes increasingly limited by $k_+$. In this
 regime, the optimal fraction of time a TF spends on the DNA increases
 with decreasing $k_+$, rising to values above $50\%$. The reason is
 that, when $k_+$ is low, the TF needs to slide multiple times over
 the promoter before it binds, and this requires a more exhaustive
 search on the DNA for a minimal search time. This redundancy is
 enhanced by lowering $k_d$, which increases the DNA occupancy.  Our
 results thus suggest that TFs spend more than 50\% of their time on
 the DNA, because that minimizes the search time when promoter
 association is reaction limited.

 Note that only in the case of a well mixed cytoplasm ($D_3\to
 \infty$, \textit{solid black line} in \fref{DNAOcc_kaP},
 \eref{kD_WM}), the time a TF spends on the DNA converges to 50\% as
 predicted by Slutsky and Mirny. For a finite cytoplasmic diffusion
 constant $D_3$, the fraction of time a TF is nonspecifically bound to
 the DNA is always lower than that in the well mixed case. As $D_3$
 decreases, the probability that after DNA dissociation a TF will
 rapidly rebind the DNA instead of diffusing into the bulk,
 increases. This increases the average number of times $M'$ a TF
 rapidly rebinds the DNA before it escapes into the bulk. Because a TF
 tends to rebind the DNA close to where it dissociated from it,
 rebindings increase the effective length $b_{\rm eff}$ of a DNA scan:
 $b_{\rm eff} = \sqrt{M'} b$. To counter the effect of rebinding (increasing
 $M'$) and to keep the effective scan length $b_{\rm eff}$ close to its
 optimal value, the rate of DNA dissociation $k_d$ has to be increased,
 so that $b$ is decreased. This lowers the fraction of time a TF
 spends on the DNA, as seen in \fref{DNAOcc_kaP}.

\begin{figure}
\centering
\includegraphics[scale=1.0]{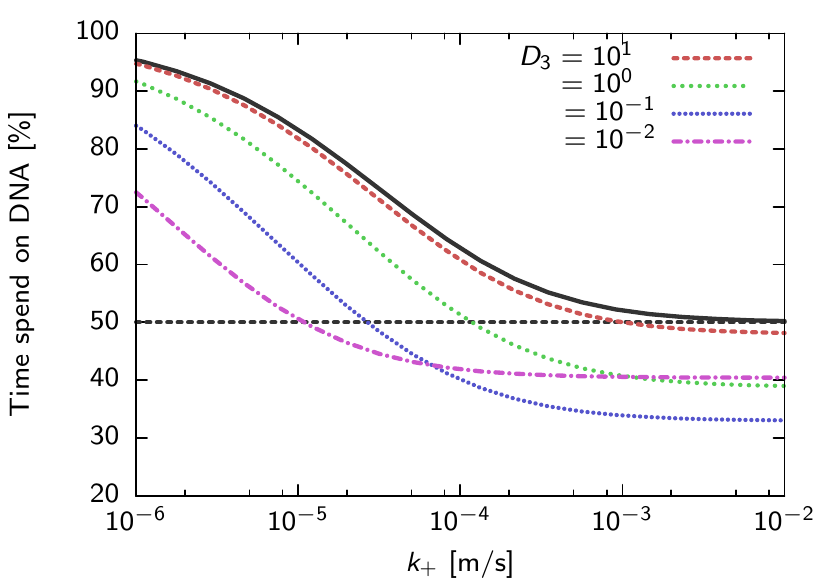}
\caption{\flabel{DNAOcc_kaP} The fraction of time a TF spends on the
  DNA as a function of the intrinsic association rate to the promoter
  $k_+$. The DNA dissociation rate $k_d$ is chosen such that it
  minimizes the search time (minimum in \fref{PromoterNoise_kdD}). In
  a well mixed cytoplasm (\textit{solid black line}), TFs spends the
  longest time on the DNA to minimize the search time, with a minimum
  of 50\% as the search becomes diffusion limited at high rates of
  $k_+$. For finite cytoplasmic diffusion constants (\textit{colored
    dashed lines}, values for $D_3$ are given in the legenda with
  units $\upmu {\rm m^2/s}$), the DNA occupancy is always lower
  because the spatial correlations of the TF with the DNA after
  dissociating from it, require a higher DNA dissociation rate $k_d$
  for an optimal search speed. Also note that in the diffusion limited
  regime (high $k_+$), the curves converge to a DNA occupancy lower
  than 50\%. Parameters: the promoter dissociation rate $k_-$ is always 
  tuned such that $\bar{n}=0.5$. Other parameters are fixed at values given in \tref{parameters}.}
\end{figure}

\subsection{CONNECTION WITH PROMOTER NOISE IN GENE EXPRESSION}
The fundamental bound on the precision of sensing TF concentrations
puts a lower bound on the contribution to the noise in gene expression
that comes from promoter-state fluctuations. Our observation that even
in the presence of 3D diffusion and 1D sliding, the promoter switches
to an excellent approximation 
in a Markovian fashion, makes it possible to quantify this
``extrinsic'' contribution.

Consider a gene which is expressed with a rate $\beta$ when a
transcription factor is bound to its promoter. The expressed protein
decays with a rate $\mu$. Our observation above shows that 
fluctuations in the promoter state $n(t)$ decay, to an excellent
approximation, exponentially with a rate $\lambda$, corresponding to a
promoter correlation time $\tau_n = \lambda^{-1}$. The noise in the
protein copy number $X$ is then given by
\cite{Paulsson2005,TanaseNicola2006}
\begin{eqnarray}
\sigma^2_X &=&\avg{X} + \left(\frac{\beta}{\mu}\right)^2
\frac{\mu}{\mu+\lambda} \sigma_n^2, \label{eq:noise_add_rule}\\
&\equiv&\sigma^2_{\rm in} + \sigma^2_{\rm ex}.
\end{eqnarray}
Here, $\sigma^2_{\rm in}=\avg{X}=(\beta / \mu) \avg{n}$ is the
intrinsic noise in $X$; it would be the noise in $X$ if the state of
the promoter were constant. The second term $\sigma^2_{\rm
  ex}$ describes the contribution to $\sigma_X^2$ from the
fluctuations in the promoter state $n$. These fluctuations are
amplified by the gain $g= \partial \avg{X}/\partial
\avg{n}=\beta / \mu $, but integrated with an integration time given by the
lifetime of the protein, $T=\mu^{-1}$ \cite{Paulsson2005}. When $T\gg
\tau_n$, we can rewrite the above expression as
\begin{equation}
\sigma^2_X = \avg{X} +  \left(\frac{\beta}{\mu}\right)^2
\frac{\tau_n \sigma_n^2}{T}.
\end{equation}
This expression highlights the idea that uncertainty in the estimate of
$\bar{n}$, $\delta n = \sigma_{n_T}$, generates fluctuations in the
expression level $X$, which are amplified by the gain $g$: $\sigma^2_{\rm
  ex}= g^2\sigma^2_{n_T}/2$. In fact, gene expression can be
interpreted as a sampling protocol, in which the history of the
promoter state $n(t)$ is stored in $X(t)$ \cite{Govern2013}. In this view, the copies of
X constitute samples of $n(t)$. This perspective
reveals that the factor 2 arises from the fact that the samples are
degraded stochastically, which effectively increases the spacing
between them \cite{Govern2013}.

\section{DISCUSSION}
We have rederived the fundamental bound on the precision of
transcriptional regulation. To this end, we have developed a theory
which is based on the model of promoter-TF binding put forward by
Tka\v{c}ik and Bialek \cite{Tkacik2009}. In this model, the DNA near
the promoter is described as a straight cylinder. This seems
reasonable since the sliding distance as measured experimentally,
$\approx 50 {\rm bp}$ \cite{Hammar2012a}, is less than the persistence
length of the DNA, which is on the order of $150 {\rm bp}$. A TF that
dissociates from the DNA, goes into the bulk where it moves by normal
diffusion at all length scales. Here, we thus ignore the interplay
between 3D diffusion, 1D sliding, hopping, and intersegmental transfer
\cite{Berg1981}. However, at length scales larger than the sliding
distance and the mesh length of the DNA polymer, the motion is
essentially 3D diffusion. At these scales, TFs move with an effective
diffusion constant, which is the result of diffusion in the cytoplasm,
hopping, intersegmental transfer and sliding along the DNA
\cite{Richter1974,Berg1981,Slutsky2004a,Hu2006,Halford2004,Elf2007,Lomholt:2009vj,Li2009,Hammar2012a}. The
diffusion constant $D_3$ in our model is indeed this diffusion
constant.

It should be realized that even in our relatively simple model,
promoter-TF binding is, in general, a complicated many-body
non-Markovian problem, because rounds of promoter-TF association and
dissociation can build up spatial-temporal correlations between the
positions of the TF molecules
\cite{Agmon1990,Kaizu2014}. Consequently, a free promoter is, in
general, not surrounded by the equilibrium distribution of TF
molecules, and the probability that a free promoter binds a TF will
depend on the history of binding events. This impedes an exact
solution of the problem.  

However, following our earlier work
\cite{Kaizu2014}, we can solve the problem almost analytically by
making one assumption and one mathematical approximation. The assumption, \eref{Def_Ssi}, is that after each
TF dissociation event, the other TFs have the equilibrium
distribution. As a result, the probability that a free promoter binds
a TF at a later time $t$, becomes independent of the history of
binding events. The approximation is that the Laplace transform of
$ \mathscr{S}_{\mathrm{rad}}(t|{\mathrm{eq}})$ is given
by \eref{Seq_approx}. The
assumption and approximation together mean that in our theory, the correlation
time of the promoter is that of a random telegraph process, where the
promoter switches between the TF bound and unbound states with rates
that are constant in time.

We have tested our theory by performing particle-based simulations of
the same model that underlies our theory. Because the GFRD algorithm
is exact and the model is the same, all deviations between theory and
simulations must be due to the assumption and/or approximation in the
theory. To test the theory, we have computed the zero-frequency limit
of the power spectrum, $P_n(\omega\to 0) = 2 \sigma^2_n \tau_n$, which
is essentially a test of the correlation time $\tau_n$, because the
variance $\sigma^2_n$  of a binary switch is
given by the mean $\bar{n}$, $\sigma^2_n = \bar{n}(1-\bar{n})$. We
find that $P_n(\omega\to 0)$ and hence the promoter correlation time
is accurately predicted by our theory.

The success of our theory is rooted in the fact that the TF
concentrations are typically low, the promoter-TF dissociation constant
is (correspondingly) low, and the sliding time is short. As a result,
the time a TF spends near the promoter is short on the timescale a TF
is specifically bound to the promoter and on the timescale new TFs
arrive from the bulk (see section \ref{sec:ValidityAssumptions}). 
A dissociated TF either rapidly rebinds the
promoter, or rapidly escapes into the bulk. This means that the
rebinding of a dissociated TF is typically not pre-empted by the
binding of another TF from the bulk---there is no rebinding
interference---which means that the central assumption of our theory,
\eref{Def_Ssi}, holds. Because TFs spend little time near the promoter
and because their concentration is low, also the mathematical
approximation, \eref{Seq_approx}, is very accurate. 

Because TFs spend only little time near the promoter, promoter
rebindings and unsuccessful bulk arrivals can be integrated out, and
the complicated many-body non-Markovian problem can be reduced to a
Markovian pair problem, in which TFs associate with and dissociate
from the promoter with effective rates that are constant in time. The
complicated dynamics of 3D diffusion and 1D sliding can thus be
captured in a well-stirred model by renormalizing the association and
dissociation rates. The off rate $k_{\rm off}$ is simply the intrinsic
dissociation rate $k_-$ divided by the average number of bindings
before escape (\eref{OffRenRates})
and the on rate $k_{\rm on}$ is the bulk arrival
rate $k_D$ times the binding probability. This probability is the
inverse of the number of bindings (\eref{OnRenRates}). This model can
then be simulated using the Gillespie algorithm
\cite{Gillespie1977,VanZon2006,Kaizu2014}. While our
model does not take into account crowding, we expect that the same
approach could be used in this case: the key observation is that inside the crowded environment of
the cell, the time a TF spends near the promoter on the DNA is short
compared to the time it is bound to the promoter and the time it takes
to arrive at the promoter from the bulk \cite{Hammar2012a}. This makes
it possible to  to study the effect of crowding on the dynamics
of gene networks using a well-stirred model  \cite{Morelli:2011ie}.

An important consequence of the fact that the promoter dynamics can,
to an excellent approximation, be described by a two-state Markov
model is that the promoter correlation time is determined by the
effective association and dissociation rates. This means that
 minimizing the sensing error, or the extrinsic noise in
gene expression, at fixed promoter occupancy corresponds to
minimizing the search time, see \eref{dNoN}.

As others have found before \cite{Berg1981,Slutsky2004a}, we find that
there exists an optimal sliding time that minimizes the search time
and hence the sensing error. Moreover, as found by Hammar et
al. \cite{Hammar2012a}, the optimal sliding distance depends on the
probability that a TF which is contact with the promoter, binds the
promoter instead of sliding over it or dissociating from the DNA into
the cytoplasm. In addition, the lower the cytoplasmic diffusion
constant, the more likely the TF will rebind the DNA after a
dissociation event, which increases the effective sliding distance. To
counteract this, the intrinsic DNA dissociation rate $k_d$ should be
increased to minimize the search time.

Finally, our model is relatively simple. For example, in our model,
the intrinsic DNA association and dissociation rates $k_a$ and $k_d$
can be changed without changing the bulk diffusion constant
$D_3$, but in reality the effective diffusion constant $D_3$ depends
on $k_a$ and $k_d$ \cite{Berg1981, Li2009, Hammar2012a}. Our results
indicate, however, that also in more realistic models of the TF
dynamics \cite{Berg1981,Li2009,Hammar2012a}, the promoter switches between the bound and
unbound states with effective rates that are constant in time. Also in
these more complex models, minimizing the sensing error will
correspond to minimizing the search time. This means that the huge
body of literature of how TFs find their target site on the DNA
\cite{Richter1974,Berg1981,Slutsky2004a,Hu2006,Halford2004,Elf2007,Lomholt:2009vj,Li2009,Hammar2012a}
can be transposed to determine the fundamental bound to the accuracy
of transcriptional regulation. Specifically, recent experiments
indicate that sliding speeds up the search process of the lac
repressor by a factor 4, compared to a hypothetical scenario where the
repressor could bind directly to the operator
\cite{Hammar2012a}. \eref{dNoN} predicts that this decreases the
fractional error in the concentration estimate by a factor of 2. 

\begin{acknowledgments}
We thank Martijn Wehrens for a critical reading of the manuscript. 
This work is part of the research programme of the Foundation for Fundamental 
Research on Matter (FOM), which is part of the Netherlands Organization for Scientific Research (NWO).
\end{acknowledgments}

\bibliographystyle{unsrt}

\appendix
\section{\alabel{noise:Seq_derived} DERIVING THE SURVIVAL PROBABILITY OF THE PROMOTER IN EQUILIBRIUM}
Here we derive the expression (\eref{DE_Seq}) for the survival
probability of a single promoter site in an equilibrated system of
transcription factors on the DNA and in the cytoplasm. Our system is
described by two coupled Green's functions; $P_1(z,t|{\rm eq})$ giving
the probability density on the DNA and $P_3(z,r,t|{\rm eq})$ the
probability density in the bulk. These functions define a single
particle problem, where the particle is initially uniformly
distributed in space. The system is finite, and cylindrically shaped
with a radius $R$ and height $L$. Inside is a rod (the DNA), lying
along the whole length of the central axis of the cylinder. By
definition, $P_1(z,t|{\rm eq})$ and $P_3(z,r,t|{\rm eq})$ are
dimensionless quantities with the following normalization
\begin{eqnarray}
 \frac{1}{L} \left. \int_{-L/2}^{L/2} dz P_1(z,t|{\rm eq})\right|_{t=0}  &=& 1 \\ \nonumber
 \left. \frac{2\pi}{V} \int_{-L/2}^{L/2} dz \int_{\sigma}^{R} rdr P_3(z,r,t|{\rm eq})\right|_{t=0} & =& 1.
\end{eqnarray}
This normalization, however, does not take into account the time the
particle spends on the DNA and in the cytoplasm. To set the proper
distribution, we introduce the dimensionless factors $\bar{P}_1$ and
$\bar{P}_3$ which give the probability of finding the particle either
on the DNA or in the cytoplasm, respectively. These probabilities have
to satisfy a detailed balance relation
\begin{equation}
 \frac{1}{L} \, \bar{P}_1 \, k_d  =  \frac{1}{V} \, \bar{P}_3 \, k_a,
 \elabel{DB_init}
\end{equation}
and have to normalize our system at $t=0$
\begin{equation}
 \frac{\bar{P}_1}{L} \int_{-L/2}^{L/2} dz P_1(z,0|{\rm eq }) + \frac{\bar{P}_3}{V} \int d\nu P_3(z,r,0|{\rm eq}) = 1.
\end{equation}
Here $\int d\nu = 2\pi \int_{-L/2}^{L/2}dz \int_{\sigma}^{R}rdr$.

We can relate this single particle problem to the survival probability of the promoter, 
surrounded by an equilibrated solution of $N$ particles, via \cite{Szabo1989} 
\begin{gather}
 \mathscr{S}_{N}(t|{\rm eq}) = \left( \frac{1}{V'} \int_{V'} d{\bf r} P({\bf r}, t|{\rm eq})\right)^N, \nonumber \\ 
 = \left( \frac{\bar{P}_1}{L} \int_{-L/2}^{L/2}dz P_1(z,t|{\rm eq }) + \frac{\bar{P}_3}{V} \int d\nu P_3(z,r,t|{\rm eq}) \right)^N,
\end{gather}
where $V'$ is the total volume of the system where particles diffuse,
including the DNA (1D) and the cytoplasm (3D). Thus, the promoter
survives, as long as none of the transcription factors in the system
have reacted. Differentiating with respect to time gives
\begin{eqnarray}
 & & \frac{1}{N \mathscr{S}_{N-1}(t|{\rm eq})} \frac{d\mathscr{S}_{N}(t|{\rm eq})}{dt} = \\ 
 & & \frac{\bar{P}_1}{L} \int_{-L/2}^{L/2}dz  \frac{dP_1(z,t|{\rm eq})}{dt} + \frac{\bar{P}_3}{V} \int d\nu \frac{dP_3(z,r,t|{\rm eq})}{dt}.\nonumber 
\end{eqnarray}
Since particles only leave the system via the promoter site positioned at $z_0$, we can 
rewrite the time derivative of the single particle problem as the radiative influx at the 
promoter site
\begin{equation}
 \frac{1}{N} \frac{d\mathscr{S}_{{\rm rad}, N}(t|{\rm eq})}{dt} = -\frac{\bar{P}_1}{L} \, k_+ P_1(z_0,t|{\rm eq})  \mathscr{S}_{N-1}(t|{\rm eq}).
\end{equation}
Taking the limit $L,N,R \to \infty$, and using that $\bar{P}_1 \frac{N}{L} = \bar{\xi}$, 
we arrive at the desired result
\begin{equation}
 \frac{d\mathscr{S}_{{\rm rad}}(t|{\rm eq})}{dt} = - \bar{\xi} k_{\rm rad}(t) \mathscr{S}(t|{\rm eq}),
\end{equation}
where $\bar{\xi}$ is the concentration on the DNA when the system is in equilibrium.

\section{\alabel{noise:rates} RELATING THE SINGLE PARTICLE SURVIVAL PROBABILITY TO REACTION RATES}
We can relate the Laplace transformed survival probability of a
promoter with only a single TF at contact, $\hat{S}_{\mathrm{rad}}(s|z_0)$,
to the intrinsic association rate of the promoter $k_+$, which is the
rate at which a TF binds the promoter given it is in contact with it,
and the (Laplace transformed) diffusion limited on-rate
$\hat{k}_{\mathrm{abs}}(s)$. The rate $k_{\mathrm{abs}}(t)$ is defined
as the rate at which TFs arrive at the promoter, starting from an
equilibrium distribution. This rate depends on the diffusion constant
in the cytoplasm, $D_3$, the diffusion constant for sliding along the
DNA, $D_1$, the rate of binding to the DNA, $k_a$, and the rate of
unbinding from the DNA into the cytoplasm, $k_d$, and the DNA
cross-section $\sigma$. The quantities $k_+$ and
$k_{\rm abs}(t)$ do not only determine $S_{\mathrm{rad}}(t|z_0)$, but
also the effective rate $k_{\mathrm{rad}}(t)$ at which TFs arrive at
the promoter and bind it.

To derive the relationship between $k_{\mathrm{rad}}(t)$,
$S_{\mathrm{rad}}(t|z_0)$, $k_+$ and $k_{\mathrm{abs}}(t)$, we exploit
the following relationships (see \cite{Agmon1990} and
\cite{Kaizu2014}). First, we note that
the time-dependent rate
constant $k_{\rm rad}(t)$ can be related to the time-dependent rate
constant $k_{\rm abs}(t)$ via
\begin{equation}
k_{\rm rad}(t) = \int_0^t dt^\prime R_{\rm rad}(t-t^\prime|z_0)
k_{\rm abs}(t^\prime),
\end{equation}
where $R_{\rm rad}(t|z_0)$ is the rate at which a TF binds the
promoter at time $t$ given it started at contact with it.  This
expression can be understood by noting that $k_{\rm abs}(t^\prime)/V$
is the probability per unit amount of time that promoter and TF
come in contact for the first time at time $t^\prime$, while $R_{\rm
  rad}(t-t^\prime)|z_0)$ is the probability that promoter and
TF which start at contact at time $t^\prime$ associate
a time $t-t^\prime$ later. In Laplace space, the above expression
reads
\begin{equation}
\hat{k}_{\rm rad}(s) = \hat{R}_{\rm rad} (s|z_0) \hat{k}_{\rm
  abs}(s).
\elabel{si:kradkabs}
\end{equation}
Since $R_{\mathrm {rad}}(t|z_0)=-\partial S_{\mathrm {rad}}(t|z_0)/\partial t$,
$\hat{R}_{\mathrm {rad}}(s|z_0)$ is also given by
\begin{equation}
\hat{R}_{\mathrm {rad}}(s|z_0) = 1 - s\hat{S}_{\mathrm {rad}}(s|z_0).
\elabel{si:RhatShat}
\end{equation}

We also know that $k_{\mathrm{rad}}(t) = k_+ S_{\mathrm{rad}}(t|z_0)$
\cite{Agmon1990}, which in Laplace space becomes:
\begin{equation}
\hat{k}_{\rm
  rad}(s) = k_+ \hat{S}_{\rm rad}(s|z_0).
\end{equation}
 Combining this with
\eref{si:kradkabs} and \eref{si:RhatShat} yields
\begin{equation}
\hat{k}_{\rm rad}(s) = \frac{k_+ \hat{k}_{\rm abs}(s)}{k_++s
  \hat{k}_{\rm abs}(s)}
\elabel{si:kradhat}
\end{equation}
and 
\begin{equation}
\hat{S}_{\mathrm{rad}}(s|z_0) =  \frac{\hat{k}_{\rm abs}(s)}{k_++s
  \hat{k}_{\rm abs}(s)}.
\elabel{si:Shat}
\end{equation}

The long-time limit of $k_{\mathrm{abs}}(t)$ is $k_{\rm D}\equiv
k_{\mathrm{abs}}(t\to \infty)=\lim_{s\to 0} s
\hat{k}_{\mathrm{abs}}(s)$. This is the rate at which TFs, which start
from an equilibrium distribution, arrive at the promoter. As mentioned
above, this rate depends on the diffusion constants in the cytoplasm
and along the DNA, $D_3$ and $D_1$ respectively, and the rates $k_a$
and $k_d$ of (non-specific) binding to the DNA. 

The long-time limit of 
$k_{\mathrm{rad}}(t)$ is $k_{\rm
  on}\equiv k_{\mathrm{rad}}(t\to \infty)=\lim_{s\to 0} s
\hat{k}_{\mathrm{rad}}(s)$. Using \eref{si:kradhat}, this yields
\begin{equation}
k_{\mathrm{on}}= \frac{k_+ k_{D}}{k_++
  k_{D}}.
\elabel{si:kon}
\end{equation}
This is the rate at which TFs, which start from the equilibrium
distribution, bind the promoter in the long-time limit. It takes into account that
not at all arrivals at the promoter lead to promoter binding. 

The long-time limit of ${S}_{\mathrm{rad}}(t|z_0)$ is
${S}_{\mathrm{rad}}(t\to\infty |z_0)=\lim_{s\to 0}
s\hat{S}_{\mathrm{rad}}(t|z_0)$, which, using \eref{si:Shat}, is
\begin{equation}
{S}_{\mathrm{rad}}(\infty|z_0) = \frac{{k}_{D}}{k_++
  {k}_{D}}.
\end{equation}

The equilibrium constant is $K_{\rm eq} \equiv k_+/k_- = k_{\rm on} / k_{\rm
  off}$. With \eref{si:kon} this yields the following expressions for
the effective association and dissociation rates:
\begin{eqnarray}
 \frac{1}{k_{\mathrm{on}}} &=& \frac{1}{k_+} + \frac{1}{k_{{D}}},\\
 \frac{1}{k_{\mathrm{off}}} &=& \frac{1}{k_-} + \frac{K_{\mathrm{eq}}}{k_{{D}}}.
\end{eqnarray}

Finally, the correlation time is given by $\tau_n = \tau_c /
{S}_{\mathrm{rad}}(\infty|z_0)$ 
  (Eq. 20 of the main text); here, $\tau_c = (k_+ \bar{\xi} + k_-)^{-1}$,
  with $\bar{\xi}$ the concentration of TFs on the DNA, is the intrinsic
  correlation time if diffusion were infinitely fast. This yields
\begin{eqnarray}
  \tau_{n} &=& \frac{k_{+} + k_{{D}}}{(k_+ \bar{\xi} + k_-)
    k_{{D}}}, \\
&=& \frac{1}{k_{\mathrm{on}} \bar{\xi} + k_{\mathrm{off}}}.
\end{eqnarray}

\nocite{Carslaw1959,Beck1992}

\end{document}